\documentclass[12pt,preprint]{aastex}
\usepackage[dvips]{epsfig}

\bibpunct{(}{)}{;}{a}{}{,}

\begin{document}
\newcommand{\hes}{HE~0557$-$4840}
\newcommand{\hen}{HE~0107$-$5240}
\newcommand{\hef}{HE~1327$-$2326}
\newcommand{\cd}{CD$-38^{\circ}\,245$}
\newcommand{\tefft}{$T_{\mbox{\scriptsize eff}}$}
\newcommand{\teffm}{T_{\mbox{\scriptsize eff}}}
\newcommand{\vradgeo}{$v_{\mbox{\scriptsize rad,geo}}$}
\newcommand{\vradhelio}{$v_{\mbox{\scriptsize rad,helio}}$}

\title{HE~0557--4840 --- ULTRA-METAL-POOR AND CARBON-RICH\footnote{Based on
observations collected at ANU's 2.3m telescope on Siding Spring Mountain,
Australia, and European Southern Observatory, Paranal, Chile (proposal
276.D--5041)}}

\author{John E. Norris\footnote{Research School of Astronomy \& Astrophysics,
The Australian National University, Mount Stromlo Observatory, Cotter Road,
Weston, ACT 2611, Australia; jen@mso.anu.edu.au},
N. Christlieb\footnote{Department of Astronomy and Space Physics, Uppsala
University, Box 515, SE-75120 Uppsala, Sweden and Hamburger Sternwarte,
University of Hamburg, Gojenbergsweg 112, D-21029 Hamburg, Germany;
norbert@astro.uu.se}, A. J. Korn\footnote{Department of Astronomy and Space
Physics, Uppsala University, Box 515, SE-75120 Uppsala, Sweden;
akorn@astro.uu.se}, K. Eriksson\footnote{Department of Astronomy and Space
Physics, Uppsala University, Box 515, SE-75120 Uppsala, Sweden;
kjell.eriksson@astro.uu.se }, M. S. Bessell\footnote{Research School of
Astronomy \& Astrophysics, The Australian National University, Mount Stromlo
Observatory, Cotter Road, Weston, ACT 2611, Australia;
bessell@mso.anu.edu.au}, Timothy C. Beers\footnote{Department of Physics and
Astronomy, CSCE: Center for the Study of Cosmic Evolution, and JINA: Joint
Institute for Nuclear Astrophysics, Michigan State University, E. Lansing, MI
48824; beers@pa.msu.edu}, L. Wisotzki\footnote{Astrophysical Institute
Potsdam, An der Sternwarte 16, D-14482 Potsdam, Germany; lutz@aip.de}, and
D. Reimers\footnote{Hamburger Sternwarte, University of Hamburg,
Gojenbergsweg 112, D-21029 Hamburg, Germany; dreimers@hs.uni-hamburg.de.}}

\begin{abstract}

\noindent We report the discovery and high-resolution, high $S/N$,
 spectroscopic analysis of the ultra-metal-poor red giant {\hes},
 which is the third most heavy-element deficient star currently known.
 Its atmospheric parameters are {\tefft}~=~4900\,K, $\log g$~=~2.2,
 and [Fe/H]~=~--4.75.  This brings the number of stars with [Fe/H] $<$
 --4.0 to three, and the discovery of {\hes} suggests that the
 metallicity distribution function of the Galactic halo does not have
 a ``gap" between [Fe/H]~=~--4.0, where several stars are known, and
 the two most metal-poor stars, at [Fe/H] $\sim$ --5.3.  {\hes} is
 carbon rich -- [C/Fe]~=~+1.6 -- a property shared by all three
 objects with ${\rm [Fe/H]} < -4.0$, suggesting that the well-known
 increase of carbon relative to iron with decreasing [Fe/H] reaches
 its logical conclusion -- ubiquitous carbon richness -- at lowest
 abundance. We also present abundances (nine) and limits (nine) for a
 further 18 elements. For species having well-measured abundances or
 strong upper limits, {\hes} is ``normal" in comparison with the bulk
 of the stellar population at [Fe/H]~$\sim$~--4.0 -- with the possible
 exception of Co.  We discuss the implications of these results for
 chemical enrichment at the earliest times, in the context of single
 (``mixing and fallback'') and two-component enrichment models.  While
 neither offers a clear solution, the latter appears closer to the
 mark.  Further data are required to determine the oxygen abundance
 and improve that of Co, and hence more strongly constrain the origin
 of this object.\\

\end{abstract}

 
\hspace {-1.0mm}{\it Suggested running title:} \hspace*{3mm} HE~0557--4840
--- ULTRA METAL-POOR AND CARBON-RICH

\section{INTRODUCTION}

The most metal-poor stars, believed to have formed at redshifts $>$ 5
\citep[see e.g.][]{Clarke/Bromm:2003}, and representing well-defined
points in space and time, hold clues on the conditions at the earliest
epochs that are provided by no other astronomical objects. That is to
say, the study of their metallicity distribution function (MDF),
together with their relative chemical abundance patterns, have the
potential to shed light on the nature of the first generation of
objects to form in the Universe, and the manner in which the
stellar-made elements (those heavier than Li) first formed.

Consider first the MDF.  Four decades after the classic analysis of
the archtypical metal-poor stars HD~19445 and HD~140283 by
\citet{Chamberlain/Aller:1951}, it could be claimed
\citep[e.g.][]{Ryan/Norris:1991} that the MDF for halo material
with\footnote{[Fe/H] = log(N(Fe)/N(H))$_{\rm{Star}}$ --
log(N(Fe)/N(H)$_{\odot}$, $\log\epsilon (\mbox{Fe})$ =
log(N(Fe)/N(H))$_{\rm{Star}}$ + 12.00} [Fe/H] $>$ --4.0 was in
reasonable agreement with the predictions of the simple closed box
model of Galactic chemical enrichment (the Simple Model) of
\citet{Searle/Sargent:1972} and \citet{Hartwick:1976}.  Following
efforts since that time, however, it has become clear that this is not
the case below [Fe/H] $<$~--4.0: while two objects are currently known
at [Fe/H] $\sim$ --5.3 (HE~0107--5840,
\citealt{HE0107_Nature,HE0107_ApJ}, and {\hef},
\citealt{Frebeletal:2005,Aokietal:2006}), the MDF presented by
\citet{Beersetal:2005a} and the Simple Model lead one to expect some
40 such objects below [Fe/H]~=~--4.0.  Figure~\ref{Fig:lowfe_MDF},
which shows the MDF for the unbiased metal-poor objects that have been
the subject of high-resolution, high signal-to-noise ($S/N$)
model-atmosphere (one-dimensional (1D), Local Thermodynamic
Equilibrium (LTE)) chemical abundance analyses\footnote{The samples of
stars identified for high-resolution analysis originate from
high-proper-motion and objective-prism surveys for metal-poor stars;
these samples are expected to suffer no metallicity-dependent bias in
their selection for [Fe/H] $<$ --2.0. The low spectral-resolution
prism surveys, which supply the vast majority of stars known with
[Fe/H] $<$ --3.0, are unable to detect the presence of the Ca~II K
line for all but the very coolest stars.  As a result, the possibility
of introducing bias into the selection function for the lowest
metallicity stars is even more remote.  When high-resolution spectra
are available one is able to guard against contamination by stellar CH
and interstellar Ca II lines that might have militated against
recognition of the most metal-poor objects in the discovery phase.
See \citet{Beers/Christlieb:2005}.}, demonstrates a second tantalizing
fact.  In contrast to the expectation of a continuously decreasing
MDF, the two objects at [Fe/H] $\sim$ --5.3 lie some 1.5 dex below the
next most metal-poor objects at [Fe/H]~=~--4.0.  Despite low numbers,
this has led some \citep[e.g.][]{Karlsson:2006} to speak of the
possibility of a metallicity ``gap" at lowest abundance, which could
have profound implications for our understanding of the nature of the
first generation of objects that chemically enriched the Galaxy, and
presumably the Universe.

Consider next the relative abundance characteristics of the two stars
with [Fe/H] $\sim$ --5.3.  Their most distinctive features are the
enormous overabundances, relative to iron, of carbon ([C/Fe] $\sim$
4), nitrogen ([N/Fe] $\sim$ 2--4), and oxygen ([O/Fe] $\sim$ 2--3)
(see also \citealt{Besselletal:2004} and \citealt{Frebeletal:2006}).
This, taken together with the well-established fact that the incidence
of carbon richness increases as one proceeds to lower abundance in the
range --4.0 $<$ [Fe/H] $<$ --2.0 \citep{Beers/Christlieb:2005},
suggests that the trend of increasing carbon-richness continues to
ubiquitous C enhancement at [Fe/H] $\sim$
--5.0. Figure~\ref{Fig:lowfe_MDF} shows this trend, where the shaded
regions represent objects with [C/Fe] $>$ 1.0.  The simplest
explanation of this phenomenon is that the earliest objects in the
Universe produced large amounts of CNO relative to Fe, far in excess
of the ratios produced at later times.  Candidate sites include
supernovae with ``mixing and fallback" (see
\citealt{Iwamotoetal:2005}, and references therein) and rotating
massive stars
\citep{Fryeretal:2001,Meynetetal:2006}. \citet{Frebeletal:2007c} argue
that relative overabundances of carbon (and/or oxygen) are necessary
below [Fe/H] = --4.0 to provide the cooling of primordial material via
\ion{C}{2} and \ion{O}{1} at the earliest times to form the low mass
stars we observe today.  We shall return to this point in \S\ref{CNO}

Two questions then beg to be addressed.  First, is the --5.3 $<$
[Fe/H] $<$ --4.0 ``gap" real, and second, what is the incidence of CNO
richness below [Fe/H] $<$ --4.0?  Given the rarity of objects below
this limit, one might expect definitive answers to be some time in
coming.  That said, the purpose of the present paper is to begin the
task.  We present here the discovery and analysis of {\hes}, a third
object having [Fe/H] $<$ --4.0.  In \S\ref{observations} we report its
discovery, and the high-resolution, high $S/N$ data obtained to
address these questions.  Our model-atmosphere analysis of this
material to produce accurate chemical abundances is then described in
\S\S\ref{parameters}--\ref{analysis}.  Finally, in \S\ref{discussion}
we discuss the significance of this object.  While much future work
remains to be done, we find that the existence of {\hes} with
[Fe/H]~=~--4.75 weakens the case for the --5.3 $<$ [Fe/H] $<$ --4.0
``gap" canvassed above. On the other hand, with [C/Fe]~=~+1.6 (and
[N/Fe] $<$ +1.5 and [O/Fe] $<$ +3.1), the data for {\hes} are
consistent with the suggestion that below [Fe/H] $<$ --4.0 all stars
possess strong enhancement of at least some elements of the CNO group.
We summarize our results in \S\ref{summary}.

\section{OBSERVATIONS}\label{observations}

\subsection{Discovery}

{\hes} was observed in 2005 December during an observing session on
candidate metal-poor stars from the Hamburg/ESO objective-prism survey
\citep[HES;][]{hespaperIII} with the Australian National University's
2.3m telescope/Double Beam Spectrograph combination on Siding Spring
Mountain.  The spectra were observed with resolving power R $\sim$
2000 and cover the wavelength range 3600--5400 {\AA}.  Comparison of
the spectrum of {\hes} with ``standard" metal-poor stars also observed
during the program immediately revealed it as an object of
considerable interest.  Figure~\ref{Fig:HE0557_lowres} shows the
spectra of {\hes} and of {\cd}, the well-known ultra-metal-poor red
giant (${\rm [Fe/H]} = -4.0$; \citealt{McWilliametal:1995b},
\citealt{Norrisetal:2001}, \citealt{Cayreletal:2004}). Comparison of
these spectra shows that the two stars have comparable hydrogen line
strength, and therefore similar effective temperature, while the Ca II
line is weaker in {\hes}, suggestive of a lower abundance of calcium.
Perhaps equally interesting is that the G band of CH at 4300 {\AA} \
is considerably stronger in {\hes}.  Given that CH is not detected in
{\cd} ([C/Fe] $<$ --0.3, \citealt{Cayreletal:2004}) it was immediately
obvious that {\hes} is carbon rich relative to {\cd}.

This object, with $\alpha (2000) = 05$\,h $58$\,m $39.2$\,s and
$\delta (2000) = -48^{\circ}\;39'\;57''$,~and $V$~=~15.45 and
$B-V$~=~0.71 (see \S\ref{photometry}), was clearly a target of
interest for high resolution follow-up spectroscopy.

\subsection{High-resolution Spectroscopy}

{\hes} was then observed in Service Mode at the Very Large Telescope
(VLT) Unit Telescope 2 (UT2) with the Ultraviolet-Visual Echelle
Spectrograph (UVES) during the nights of 2006 February 5, 25, and 26
and March 19. Seven individual exposures with an integration time of
1\,h each were obtained. UVES was used in dichroic mode with the
BLUE390 and RED580 settings. The useful wavelength ranges, in the rest
frame of {\hes}, are 3300--4520\,{\AA} in the blue-arm spectra, and
4820--5750\,{\AA} and 5835--6800\,{\AA} in the lower- and upper-
red-arm spectra, respectively.

A $1''$ slit was used in both arms, yielding a nominal resolving power
of $R\sim 40,000$. Since read-out noise was not a limitation, $1\times
1$ pixel binning was employed during the observations, even though
this results in a significant oversampling due to the scale of
$0.215''$ per pixel of the blue-arm CCD and $0.155''$ per pixel in the
red-arm.

The seven pipeline-reduced spectra were co-added in an iterative
procedure in which pixels in the individual spectra affected by cosmic
ray hits not fully removed during the data reduction, affected by CCD
defects, or other artifacts, were identified. These pixels were
flagged and ignored in the final iteration of the co-addition.

The co-added blue-arm spectra, rebinned by a factor of 2, have a
maximum $S/N$ per pixel of $120$ at $4200$\,{\AA}, decreasing to $S/N
\sim 100$ at the red end. At the blue end, this decreases to
$S/N~=~60$ at 3700\,{\AA}, and $S/N~=~15$ at 3300\,{\AA}. In the
rebinned lower red-arm spectra, the $S/N$ increases linearly from
$160$ per pixel at 4820\,{\AA} to $200$ at 5750\,{\AA}. It is
approximately constant at a level of $S/N~=~200$ per pixel throughout
the wavelength range covered by the upper red-arm spectra.

\subsubsection {Line Strength Measurement}

Equivalent widths for {\hes} were measured independently by the first
two authors, using techniques described by \citet{Norrisetal:2001} and
\citet{HE0107_ApJ}, for unblended lines defined in those
investigations, together with that of \citet{Cayreletal:2004}. A
comparison of the two data sets is presented in
Figure~\ref{Fig:EqwTestHE0557}, where the agreement is quite
satisfactory, with an RMS scatter about the line of best fit of 2.5
m{\AA}.  In the absence of an understanding of the origin of the small
departure from the one-to-one line evident in the figure, we have
chosen to average the data, and present results for 100 lines suitable
for abundance analysis in column (5) of Table~\ref{Tab:Linelist}.  The
atomic data in the table were mostly taken from the Vienna Atomic Line
Database (VALD\footnote{\texttt{http://www.astro.uu.se/$\sim$vald/}};
\citealt{VALD2a}, \citealt{VALD2b}), complemented where necessary by
values from the three works cited above. For completeness, we identify
in Table~\ref{Tab:RejectedLines} of Appendix A a further 45 lines that
we regarded as unsuitable for abundance analysis.

\subsubsection{Radial Velocities}\label{radial_velocities}

Heliocentric radial velocities for {\hes} were measured for the seven
individual spectra described above by Gaussian fits of six clean and
moderately strong absorption lines.  The resulting values are
presented in Table~\ref{Tab:HE0557_Velocities} where columns (1) and
(2) contain the Modified Julian Date and heliocentric velocity,
respectively.  The mean internal standard error of measurement of a
single observation is 0.2 kms$^{-1}$, while the external error is
estimated to be 0.7 kms$^{-1}$ \citep{Napiwotzkietal:2007}.

Given that the dispersion of velocities in the table is 0.21
kms$^{-1}$, we conclude that to within the accuracy of our
measurements the velocity of {\hes} did not vary over the 43 days of
the present observations.

It is interesting to note the large radial velocity, 211.8 kms$^{-1}$,
of {\hes}, in comparison with values of 44.1~kms$^{-1}$ for {\hen}
\citep{HE0107_ApJ} and 63.8~kms$^{-1}$ for {\hef}
\citep{Aokietal:2006}.  The large spread in these three values
establishes that stars with [Fe/H] $<$ --4.0 have a velocity
dispersion commensurate with that of more metal-rich halo material
([Fe/H] $>$ --4.0).

\subsection{Photometry}\label{photometry}

In Table~\ref{Tab:HE0557_Photometry}, we present photometry for
{\hes}. The values for the $BVR$ photometry were taken from
\citet{Beersetal:2007}, while the $K$ magnitudes are from
\citet{Skrutskieetal:2006}. In this table we also list data for
HE~0107--5840 ([Fe/H]~=~--5.3) from \citet{HE0107_ApJ}, for comparison
purposes. For the reddening of {\hes} we adopt $E(B-V)$ = 0.04
following \citet{Schlegeletal:1998}, and the differential
relationships $E(V-R)~=~0.62E(B-V)$ and $E(V-K)~=~3.07E(B-V)$
following \citet{Besselletal:1998}. For {\hen} we use
$E(B-V)$~=~0.013.

\section{ATMOSPHERIC PARAMETERS}\label{parameters}

\subsection{Effective Temperature}

Following \citet{HE0107_ApJ}, we employ the calibrations of
\citet{Alonsoetal:1999b,Alonsoetal:2001} and
\citet{Houdasheltetal:2000} to determine effective temperatures for
both {\hes} and {\hen}. Our results are presented in
Table~\ref{Tab:HE0557_Teff}, where the errors correspond to
uncertainties of 0.02, 0.02, and 0.06\,mag in $(B-V)_{0}$,
$(V-R)_{0}$, and $(V-K)_{0}$, respectively, based on the errors in
photometry cited by \citet{Beersetal:2007} and \citet{HE0107_ApJ}, and
reasonable errors in reddening.  Weighting the values from the three
colors in the table by the inverse square of their errors, we find
photometric temperatures $\teffm = 5090$\,K and $5170$\,K for {\hes}
and {\hen}, respectively. The present results for {\hen} agree well
with those of \citet{HE0107_ApJ}.

We note that the $(B-V)$ and $(V-R)$ colors indicate insignificant
effective-temperature differences between {\hes} and {\hen}, while the
two calibrations of $(V-K)$ point towards an effective-temperature
difference of around $300$\,K between the two objects. The source of
this difference is not understood.  For completeness, we also note
that for {\hes} $\teffm(V-K)$ $-$ $\teffm(B-V,V-R)$ $\sim$ $-150$\,K,
while for {\hen} the corresponding quantity is $+170$\,K.

To constrain the effective temperature further, we compare the
observed profiles of H$\alpha$ and H$\beta$ with synthetic profiles
based on the recipes described in \citet{Fuhrmannetal:1993}. While
this implementation does not take into account the latest advances in
the modelling of the self-broadening of Balmer lines
\citep{Barklemetal:2000a}, the use of that theory would lower the
effective temperature by no more than 50\,K, as is evident from the
analysis of {\hen} \citep{HE0107_ApJ}. Even smaller differences are
obtained when the effective temperatures are determined relative to
the \tefft\ value derived for the solar Balmer lines in a differential
analysis.  H$\alpha$ (see Figure~\ref{Fig:Halpha}) and H$\beta$
unanimously point towards an effective temperature of 4900 $\pm$
100\,K, lending support to the photometric calibrations giving a large
($\sim$ $300$\,K) effective-temperature difference between {\hes} and
{\hen} (see column (6) of Table~\ref{Tab:HE0557_Teff}). While the line
strength of H$\alpha$ directly correlates with gravity, the opposite
is true for H$\beta$ and the higher-order Balmer lines. In addition to
determining $\teffm$, the combined analysis of H$\alpha$ and H$\beta$
allows one to set constraints on $\log g$ as well (see \S\ref{logg}).

Given this difference in photometric and spectroscopic effective
temperatures, in what follows we shall present abundances for {\hes}
determined for both $\teffm~=~4900$\,K and $5100$\,K.

\subsection{Microturbulence and Iron Abundance}\label{micro}

Using 1D model atmospheres described in \S\ref{atmospheres} below and
assuming LTE, we determined the microturbulence velocity $\xi_{\rm
micr}$ from 60 \ion{Fe}{1} lines by requiring that there be no trend
of Fe abundance with line strength.  All \ion{Fe}{1} lines below
3700\,{\AA} were ignored, because the $S/N$ drops rapidly towards
shorter wavelengths, resulting in a higher uncertainty of the
equivalent width measurements and an increased line-to-line scatter of
the \ion{Fe}{1} abundance of up to 0.8\,dex. The average \ion{Fe}{1}
abundances are $\log\epsilon~=~2.77$\,dex and $3.02$\,dex for $\teffm
= 4900$\,K and $5100$\,K, respectively \footnote{ We note for
completeness that this procedure does not yield abundances completely
independent of excitation potential.  Inspection of
Table~\ref{Tab:Linelist}, for example, reveals that for $\teffm =
4900$\,K $d(\log\epsilon/d\chi~\sim~-0.2$\,dex/eV.  We have
encountered this effect in our analyses of other metal-poor giants,
and acknowledge it as a shortcoming of the present 1D analysis.}.  In
both cases, $\xi_{\rm micr}$ = 1.8 kms$^{-1}$.

We also detect two very weak lines of \ion{Fe}{2} in our red-arm
VLT/UVES spectra. The averaged equivalent widths are 4\,m{\AA} for the
line at $5018.44$\,{\AA} and 8\,m{\AA} for the line at
$5169.03$\,{\AA}. For the latter line, however, spectrum synthesis has
to be used to derive an abundance, because the line is blended with a
weak \ion{Fe}{1} line at $5168.90$\,{\AA}, having excitation potential
$\chi~=~0.052$\,eV and $\log gf~=~-3.97$ according to VALD. Taking the
blend into account and using atomic data from VALD (see
Table~\ref{Tab:Linelist}), the abundances inferred from the two
\ion{Fe}{2} lines are discrepant by $\sim 0.3$\,dex for
$\teffm~=~4900$\,K.  This difference can be partly removed by using
``astrophysical'' $\log gf$ values for the two lines.  That is to say,
the $\log gf$ values are determined by fitting the solar spectrum,
resulting in values of $\log gf~=~-1.40$ and $-1.30$ for \ion{Fe}{2}
$5018.44$ and $5169.03$, respectively, where following
\citet{Asplundetal:2005a} we adopt $\log\epsilon (\mbox{Fe})$ =
7.45. The abundance difference is then reduced to $0.21$\,dex.

There is also considerable uncertainty in the measurement of the
equivalent widths of these lines, because they are weak, and therefore
small changes in the continuum placement have a large effect on the
measured line strength.  Furthermore, usage of different fitting
methods (e.g., using a fixed line width when fitting a Gaussian
profile to the line as opposed to leaving the line width as a free
parameter during the fit) result in measurements deviating
significantly from each other.  The uncertainty is illustrated by the
fact that the measurements of the $5018.44$\,{\AA} line by
A.J.K. (6.6\,m{\AA}) and N.C. (4.4\,m{\AA}) differ by 2.2\,m{\AA},
translating to an abundance difference of
$\Delta\log\epsilon~=~0.18$\,dex.

In summary, the abundance difference of $0.3$\,dex can be
well-explained by the combined uncertainties of the line strength
measurements and $\log gf$ values for the two lines.

The iron abundance determined from \ion{Fe}{2} lines is less sensitive
to changes of {\tefft} than that from \ion{Fe}{1} lines, and therefore
the former is less affected by possible systematic errors in the
adopted {\tefft} scale.  Further, \ion{Fe}{2} lines are less affected
by non-LTE effects than those of \ion{Fe}{1}.  For these reasons, we
adopt the abundance determined from the two \ion{Fe}{2} lines as the
Fe abundance of {\hes}.

\subsection{Surface Gravity}\label{logg}

Establishing the ionization equilibrium of an element is a standard
spectroscopic method of deriving the stellar surface gravity. To
constrain the surface gravity of {\hes}, we perform a non-LTE analysis
using the iron model atom and line data calibrated on local halo stars
with good HIPPARCOS parallaxes \citep{Kornetal:2003}. With a mean
overionization of $+0.11$\,dex in Fe I, the ionization equilibrium is
balanced at $4900$\,K (from H$\alpha$ and H$\beta$) and $\log
g$~=~2.2. Assuming an effective temperature of $5100$\,K, the iron
ionization equilibrium would require a much higher gravity, of the
order of $\log g$ = 3. Such a stellar-parameter combination is in
conflict with both the gravity derived from an appropriate isochrone
-- following the procedure of \citet {HE0107_ApJ} for {\hen}, one
obtains log$g$ = 2.2 -- and the constraints of Balmer-profile fitting
(see Figure~\ref{Fig:Halpha}).  As we shall see in \S\ref{calcium} the
Ca~I/Ca~II ionization equilibrium at {\tefft}~=~4900\,K is also
consistent with $\log g$~=~2.2.

For various reasons (use of a slightly different model atmosphere and
$gf$ values, different line selection etc.) the LTE ionization
equilibrium was established by N.C. at $\log g$~=~$2.4$; that is, at a
higher gravity than indicated by the non-LTE analysis, but without any
consideration of overionization. The combination of Balmer-profile
temperatures and Fe I/Fe II non-LTE surface gravities was recently
shown to yield stellar parameters for metal-poor giant stars in
excellent agreement with stellar evolution
\citep{Kornetal:2006}. Nonetheless, given the uncertainties arising
from the unknown absolute temperature scale of metal-poor stars, we
have to accept a potential systematic error in absolute $\log g$
values of up to 0.3\,dex.

\subsection{Summary}

Table~\ref{Tab:StellarParameters} summarizes our atmospheric
parameters for {\hes}, based on spectroscopic and photometric data.
While we consider the purely spectroscopic stellar parameters to be
more trustworthy, abundance ratios for both solutions are presented
for the reader's convenience in what follows.  The iron abundance for
{\hes} is [Fe/H] = --4.75 or --4.71, depending on which of the
spectroscopic or photometric solutions, respectively, one prefers.

\section{MODEL ATMOSPHERES}\label{atmospheres}

We have computed 1D model atmospheres tailored for the abundances of
{\hes}; i.e., taking into account in particular the over-abundance of
C, as determined during a next-to-final iteration of our abundance
analysis. An enhancement of the $\alpha$-elements, including oxygen,
of $+0.4$\,dex, was assumed.

The models were computed with the MARCS code, using opacity sampling
in about 100,000 wavelength points. Atomic and diatomic molecular
opacities were included in addition to the continuous ones in the
model atmosphere computations. The models are hydrostatic, spherically
symmetric and computed under the assumption of LTE; convective energy
transport is included using a Mixing Length Theory
formulation. Further details of the model atmospheres can be found in
\citet{Gustafssonetal:2007}.

\section{ABUNDANCE ANALYSIS}\label{analysis}

\subsection {Atomic features}\label{Atomic}

Following \citet{HE0107_ApJ}, we computed LTE abundances for the
transitions in Table~\ref{Tab:Linelist}, using the atomic data and
equivalent widths (and their limits) presented there, the atmospheric
parameters of Table~\ref{Tab:StellarParameters}, and the code
\texttt{eqwi}, version 7.04.  The results are presented in columns (6)
and (7) of Table~\ref{Tab:Linelist}, for $\teffm~=~4900$\,K and
$5100$\,K, respectively.  For a few lines (in particular those for
which only an upper line strength limit appears in Table
\ref{Tab:Linelist}) spectrum synthesis was performed with the code
\texttt{bsyn}, version 7.04.  Figure~\ref{Fig:SrII4077synth} presents
an example of spectrum synthesis in the region of the Sr II 4077.71
line, which was used to confirm the upper limit to the strontium
abundance derived by the equivalent width analysis.  Spectrum
synthesis was also used in the analysis of lines of \ion{Li}{1},
\ion{Ca}{1}, \ion{Ca}{2}, and \ion{Fe}{2} (see \S\ref{micro}).

\ion{Co}{1}, with numerous weak lines in the range
$\sim$~3400--3600\,{\AA} (and hence difficult to measure in {\hes},
given the $S/N$ of the spectra), was analysed by using spectrum
co-addition for several lines as described by \citet{Frebeletal:2006}
in their analysis of OH lines in HE~1300+0157.  (In this method
several lines are chosen, the wavelength scale is set to zero at line
centre for each line, and the spectra co-added to yield a composite
spectrum.)  Our adopted application of the technique is shown for the
co-addition of the strongest two \ion{Co}{1} lines in
Figure~\ref{Fig:CoaddCo}, which presents a comparison of observed and
synthetic spectra of {\hes} and {\cd}.  (For {\hes} the synthetic
spectra pertain to {\tefft}~=~4900\,K, while for {\cd} we adopt
$\teffm~=~4900$\,K, log$g$~=~2.0 following \citet{Besselletal:2004},
$\xi_{\rm micr}$~=~2.4 kms$^{-1}$, $\log\epsilon (\mbox{\ion{Fe}{2}})$
~=~3.39 from \citet{Christliebetal:2007}, and $\log\epsilon
(\mbox{Co})$~=~1.44, and hence [Co/Fe]~=~0.58 obtained for the two
co-added lines (Christlieb, priv. comm.)).  For {\hes} and {\cd}
Figure~\ref{Fig:CoaddCo} yields [Co/Fe] = 0.04 and 0.7,
respectively, with the latter in good accord with the result of
Christlieb et al. cited above.

Our final results for the atomic lines of 18 elements (ten detections,
eight upper limits) are presented in Table~\ref{Tab:1DLTEAbundances}.
  
\subsection{Molecules}

Assuming LTE, we have used spectrum synthesis techniques to seek to analyze
features of CH, NH, and CN and constrain the abundances of carbon,
nitrogen, and oxygen.
 
\subsubsection{Carbon}

The C abundance of {\hes} was determined from weak to moderately
strong CH A-X lines in the wavelength region 4252--4256\,{\AA} (see
Figure \ref{Fig:CH4250fit}), the head of the G band at $\sim
4310$\,{\AA}, the feature at $\sim 4323$\,{\AA} (see
Figure~\ref{Fig:CH4323fit}), and weak lines of the B-X system in the
region around \ion{Ca}{2}~K. We used a line list compiled by B. Plez
and A. Jorisson (2006, priv. comm.). The $\log gf$ values and line
positions were taken from LIFBASE \citep{Luque/Crosley:1999}, and the
excitation energies from \citet{Jorgensenetal:1996}. The abundances
all agree to within 0.05\,dex with each other, and we adopt the
average of the results, which is listed for both combinations of the
stellar parameters in Table \ref{Tab:1DLTEAbundances}.

\subsubsection{Nitrogen}

We determined an upper limit for the nitrogen abundance of {\hes} by
analyzing the $(0,0)$ band head of the B-X system of CN at 3883\,{\AA}
(see Figure \ref{Fig:CN3883synth}). In these calculations we use a
line list of B. Plez (2001, priv. comm.). The details of its
computation are described in \citet{Hilletal:2002}. We adopt a
dissociation energy for the CN molecule of $D_{0}^{0}~=~7.76$\,eV, and
the C abundance obtained from the CH lines.

Stronger upper limits can be inferred from the absence of the $(0,0)$
band head of the A-X system of NH at $\sim 3360$\,{\AA} (see Figure
\ref{Fig:NHsynth}), yielding $\log\epsilon\left({\rm N}\right) < 4.50$
and $<5.00$ for $\teffm~=~4900$\,K and $\teffm~=~5100$\,K,
respectively. For our spectrum synthesis calculations we used the line
list of \citet{Kurucz:2006} with the $\log gf$ values reduced by
$0.3$, and a dissociation energy for NH of $D_{0}^{0}~=~3.45$\,eV. For
a discussion of the reasons for the correction of the $\log gf$ values
and the choice of $D_{0}^{0}$ see \citet{Johnsonetal:2007}.

\subsubsection{Oxygen}

In Table~\ref{Tab:1DLTEAbundances}, we have listed upper limits for
the oxygen abundance, derived from the upper limit to the line
strength of the \ion{O}{1} line at 6300.30\,{\AA}. Due to our VLT/UVES
spectrum beginning only at 3300\,{\AA} it was not possible to
determine the oxygen abundance from the UV OH A-X lines.

\subsection{Calcium in non-LTE}\label{calcium}

As well as iron, calcium is also observed in two ionization
stages. This offers the possibility of checking the spectroscopic
stellar-parameter determination, primarily the gravity,
independently. As can be seen from the three calcium-line entries of
Table~\ref{Tab:Linelist}, the two ionization stages indicate
significantly different LTE abundances.  For the {\tefft}~=~4900\,K
case, $\log \epsilon$ (Ca I 4226)~=~$1.81$ and $\log \epsilon$ (Ca II
3933) = $2.26$. The abundance derived from the subordinate line Ca II
3706 falls in between -- $\log \epsilon$ (Ca II 3706)~=~$2.11$ with
some added uncertainty arising from blends with CH. The situation is
similar for the {\tefft}~=~5100\,K solution, albeit at higher calcium
abundances.

According to computations by \citet{Mashonkinaetal:2007}, both Ca I
4226 and Ca II 3706 are affected by non-LTE. Such corrections raise
the abundances derived from these lines by $\sim$ +0.30 and
+0.17\,dex, respectively. The non-LTE corrections are similar for both
solutions. While this brings the two observed Ca II lines into
excellent agreement for the {\tefft}~=~4900\,K solution, the predicted
non-LTE effect for Ca I 4226 is too small to establish the ionization
equilibrium (by 0.2\,dex). Thus, while the two lines of Ca II can be
reconciled at 4900\,K, the ionization equilibrium seems to ask for a
significantly lower surface gravity than log$g$~=~2.2.

For the {\tefft}~=~5100\,K solution, the non-LTE abundance of Ca I
4226 is 0.07\,dex larger than that derived from Ca II 3706. This
indicates that the gravity would have to be raised by about 0.3\,dex
to establish the ionization equilibrium.  At these stellar parameters,
Ca II 3933 would yield an abundance 0.18\,dex above the other two
lines. The corresponding trend in abundance versus excitation energy
in Ca II is reminiscent of similar trends seen in Fe I for metal-poor
stars including \hes\ and could indicate deficits in the T$-\tau$
relation of hydrostatic model atmospheres.

Table \ref{Tab:calcium} indicates how the situation would change, for
{\tefft}~=~4900\,K, if the analysis were done differentially with
respect to the Sun employing the methodology and $gf$ values favored
by \citet{Mashonkinaetal:2007}. Column (2) presents the non-LTE
corrections discussed above, while columns (3) and (4) give the
changes due to different $gf$ values, and fitting to the Sun,
respectively.  (Since both lines of Ca II are very strong and heavily
blended in the solar spectrum, no attempt was made to derive
astrophysical $gf$ values for them.  In these cases, only the
corrections given in columns (2) and (3) were employed.)  The final
calcium abundances are presented in column (5).

As can be seen, the ionization equilibrium of Ca I 4226 and the $\log
g$-sensitive line Ca II 3706 can be established to within 0.03\,dex at
{\tefft}~=~4900\,K. The three lines then indicate a calcium abundance
$\log \epsilon$ (Ca) = $2.22$~$\pm$~0.03 (1$\sigma$). For the
{\tefft}~=~5100\,K solution, the abundance from Ca I 4226 is 0.25\,dex
larger than that from Ca II 3706. This would require a substantially
($\sim$ 1\,dex) larger $\log g$ (cf. \S\ref{logg}). Accordingly, the
abundances from the three lines scatter significantly more -- $\log
\epsilon$ (Ca) = $2.39$~$\pm$~0.14, if $\log\,g$\,=\,2.2 is assumed.

We conclude that, within the uncertainties of the present analysis, it
is possible to establish the Fe I/Fe II and Ca I/Ca II ionization
equilibria at {\tefft}~=~4900\,K and $\log g$~=~2.2, while this is not
the case for {\tefft}~=~5100\,K and $\log g$~=~2.2. Adding
Balmer-profile analyses to the picture, we therefore favour the
{\tefft}~=~4900\,K solution. We note that the chosen stellar-parameter
determination thus employs techniques very similar to those used in
the analysis of \hen\ \citep{HE0107_ApJ}, putting the chemical
abundances of both stars on the same abundance scale and making
possible a differential analysis.

\subsection{Abundance Errors}

Abundance errors, within the 1D, LTE formalism adopted here, have been
discussed in considerable detail by \citet{HE0107_ApJ} for {\hen}, to
which we refer the reader. Given the similarity of temperatures and
gravities of that object and {\hes}, we shall not repeat the
discussion here, except to note that random abundance errors
introduced by uncertainties of the measurements of equivalent widths
in a spectrum of the quality that we use here are typically of the
order of $0.1$\,dex, and those caused by uncertainties of atomic data
are typically of the order of $0.10$--$0.15$\,dex. Systematic errors
of the stellar parameters, most notably {\tefft}, can easily result in
systematic difference in the \emph{absolute} abundances of 0.2\,dex,
while these errors mostly cancel out for abundance \emph{ratios} of
atomic species of the same ionization state having similar ionization
potentials (e.g., [\ion{Ti}{2}/\ion{Fe}{2}], or
[\ion{Ni}{1}/\ion{Fe}{1}])

Non-LTE corrections are metallicity dependent and can become quite
large in extremely metal-poor stars (especially giants), in particular
for resonance lines of photoionization-dominated minority
species. This is, for example, the case for Ca I 4226 for which the
non-LTE correction is $\sim$ +0.5\,dex for \hef. Smaller corrections
are expected for lines arising from transitions within dominant
ionization stages.  However, for subordinate lines corrections can
reach $\sim$ 0.2\,dex (cf. Ca II 3706 as discussed in
\S\ref{calcium}).

Finally, in extremely metal-poor giants, the work of
\citet{Colletetal:2006} shows that the use of 3D, LTE modeling (as
opposed to our 1D, LTE procedure) will lead to significantly lower
abundances than reported here.  In column (6) of
Table~\ref{Tab:1DLTEAbundances} we present their results for the
corrections $\Delta\log\epsilon (\mbox{X})$ = $\log\epsilon
(\mbox{X})_{\rm 3D}$ -- $\log\epsilon (\mbox{X})_{\rm 1D}$ for {\hen}
(which has very similar {\tefft} and $\log g$ to those of {\hes}).
Roughly speaking, our abundances derived from diatomic molecules are
too high by $\sim$~1.0~dex, while for atomic species our values are
too large by --0.1 to 0.5, with most values in the range 0.1 -- 0.3.
That said, one should note the caveat of \citet{Colletetal:2006} in
the context of the iron peak elements: ``We caution ... that the
neglected non-LTE effects might actually be substantial ..."

\section{DISCUSSION}\label{discussion}

While Table~\ref{Tab:1DLTEAbundances} presents abundances for both
$\teffm~=~4900$\,K and $\teffm~=~5100$\,K, for reasons stated above,
our preference is for the former.  Accordingly, we shall use results
for $\teffm~=~4900$\,K in what follows.

\subsection{The Low-metallicity Tail of the Metallicity Distribution Function}  

With [Fe/H]~=~--4.75, {\hes} falls midway between the most metal-poor
objects at [Fe/H] $\sim$~--5.3 ({\hen} and {\hef}), on the one hand,
and the several next metal-poor stars at [Fe/H] --4.0, on the other.
{\it Our first conclusion is that the discovery of {\hes} weakens the
case for a ``gap" in the metallicity distribution function (MDF)
between [Fe/H]~=~--4.0 and --5.3}.

We also note that although the search for metal-poor stars in the HES
is not yet complete, the survey has already increased the total number
of known metal-poor stars by more than a factor three with respect to
all previous surveys combined for such objects, and all three stars
currently known to have ${\rm [Fe/H]} < -4.0$ were discovered in the
HES. This suggests that the previously suspected absence of stars at
${\rm [Fe/H]} < -4.0$, as well as the presence of the putative ``gap"
in the MDF were statistical artifacts caused by too small samples of
metal-poor stars. If this explanation is correct, it is expected that
the follow-up observations of the remaining HES metal-poor candidates,
and even deeper surveys, such as SEGUE \citep[see
e.g. \texttt{http://www.sdss.org/segue} and the description
in][]{Beers/Christlieb:2005}, will further populate the ``gap'', as
well as the metallicity region below ${\rm [Fe/H]} = -5.0$.

The shape of the low-metallicity tail of the MDF of the Galactic halo
contains information on the earliest phases of its chemical
enrichment.  \citet{Prantzos:2003}, for example, compares a
preliminary MDF derived from the results of spectroscopic follow-up
observations of candidate metal-poor stars from the HK survey (see
e.g. \citet{Beersetal:2007} for details of the survey) with three
different models of galactic chemical evolution. He finds that the
low-metallicity tail is reproduced well by a ``simple outflow model''
with a relaxed Instantaneous Recycling Approximation (IRA) and the
assumption of an early phase of infall of primordial gas lasting about
0.2\,Gyr. The latter assumption reduces the number of stars at
$\mathrm{[Fe/H]} < -3.0$ relative to ``simple outflow'' models without
early infall and with or without IRA. Such a reduction is adopted to
match the observed MDF. The model predicts, however, an MDF increasing
continuously from the lowest [Fe/H] to the peak of the MDF at around
$\mathrm{[Fe/H]}=-1.6$.  From Figure 1 of \citet{Prantzos:2003}, one
sees that the number of stars in the range $-5.0 < \mathrm{[Fe/H]} <
-4.0$ should be about five times as high as the number having
$\mathrm{[Fe/H]} < -5.0$. Since there are now two stars known with
$\mathrm{[Fe/H]} < -5.0$, about ten stars should have been found in
the previously suspected ``gap'' of the MDF -- while in reality only
one has been identified ({\hes}).

Similar problems arise for the galaxy formation and chemical evolution
model of \citet{Salvadorietal:2007}. A critical metallicity
$Z_{\mathrm{cr}}$ for low-mass (i.e., $M < 1\,M_{\odot}$) star
formation as low as $Z_{\mathrm{cr}} = 10^{-5}\,Z_{\odot}$ would have
to be adopted in order to reproduce the number of stars at
$\mathrm{[Fe/H]} < -5.0$ (i.e., two) correctly, but this would result
in a large number of objects in the regime $-5.0 < \mathrm{[Fe/H]} <
-4.0$, in contradiction with the observations.

\citet{Karlsson:2006} considered a stochastic chemical enrichment
model for the Galactic halo in which negative feedback effects, from
the first generation of (massive) stars that formed from primordial
gas clouds, resulted in a suppression of all star formation.  Later,
at redshifts of about $z=5$, the second generation of stars formed
from gas pre-enriched to levels of $\mathrm{[Fe/H]} < -5.0$ by the
previous generation of stars that exploded as Type-II supernovae
(SN~II).  (The resulting mass function had a bias towards high mass
(short-lived) stars except where carbon was sufficiently overabundant
with respect to iron to allow low mass (long-lived) objects to form.)
The MDF resulting from this model has one (or more) ``pre-enrichment''
peaks in the range $-5.5 < \mathrm{[Fe/H]} < -5.0$, with the number of
stars then first decreasing with increasing [Fe/H], until it rises
again at $\mathrm{[Fe/H]} > -4.0$.

The MDF of Karlsson matches the observations reasonably well.  In his
 most favourable case, with suppression of star formation in
 carbon-deficient gas, some 3--4 objects are predicted in the range
 --5.1 $<$ [Fe/H] $<$ --4.2.  The observational constraints on the
 exact form of the MDF at the lowest metallicities are, however, still
 weak, since currently only three stars having $\mathrm{[Fe/H]} <
 -4.0$ are known. Ongoing and future surveys for metal-poor stars,
 expected to lead to the identification of significant numbers of new
 stars at the lowest metallicities as noted above, should clarify the
 situation.

A particularly interesting and exciting prediction of Karlsson's model
is that there is an additional, smaller peak in the MDF at around
$\mathrm{[Fe/H]} = -7.0$. From his Figure 2, one can estimate that
among ten stars at $\mathrm{[Fe/H]} < -5.0$ there should be one at
$\mathrm{[Fe/H]} \sim -7.0$\footnote{One should also bear in mind the
possible role of accretion onto the most metal-poor stars of more
``normal" abundance material from the Galactic ISM over some 13\,Gyr,
as discussed, for example, by \citet{Yoshii:1981} and
\citet{Iben:1983}.  The latter predicted that such effects could raise
the iron abundance in the outer layers of red giants from zero to
[Fe/H] = --5.7.  We refer the reader also to the discussion by
\citet{HE0107_ApJ} who considered possible maximal accretion onto
{\hen}, and concluded that ``most heavy elements of {\hen} {\it might}
be accreted from the interstellar medium, although this is most
probably not the case for C, N, and Na".  What remains to be addressed
is whether minimal accretion conditions existed that would permit one
to recognize objects that initially had [Fe/H] = --7.0.}

\subsection{Relative Abundances}\label{RelativeAbundances}

Figure~\ref{Fig:Relative_Abundances} presents [X/Fe] as a function of
[Fe/H] for 12 representative elements (X) in {\hes} and the other
$\sim$~50 metal-poor stars with [Fe/H] $<$ --3.0 for which
high-resolution, high $S/N$ abundance analyses exist.  The comparison
data have been taken with first preference from
\citet{Aokietal:2002d,Aokietal:2004a,Aokietal:2006},
\citet{Besselletal:2004}, \citet{KeckpaperII},
\citet{Cayreletal:2004}, \citet{HE0107_ApJ}, \citet{KeckpaperIV},
\citet{Francoisetal:2003}, \citet{Frebeletal:2007a},
\citet{Norrisetal:1997b,Norrisetal:2000,Norrisetal:2001,Norrisetal:2002},
and \citet{Plez/Cohen:2005}, supplemented by earlier material from
\citet{McWilliametal:1995b} and \citet{Ryanetal:1991,Ryanetal:1996}.
In the figure, {\hes} is represented by a filled star, HE~1300+0157
(Frebel et al.)  by an open one, results of Cayrel et al. and Fran{\c
c}ois et al.\ by filled circles, and others by open ones.

It is important to note that all of the data in the figure have been
determined using 1D model atmospheres and the assumption of LTE.  This
should be borne in mind when comparison is made with predictions of
stellar evolution and galactic chemical enrichment models.

We note in passing that in Figure~\ref{Fig:Relative_Abundances} there
are six pairs of related elements -- representing the CNO group, the
light odd-numbered elements, the $\alpha$-elements, the Fe-peak below
(Cr and Mn) and above (Co and Ni) iron, and the heavy neutron-capture
elements.
 
As has been discussed at some length in a number of the works cited
here, and is clear in the figure, the case can be made that for [Fe/H]
$<$ --3.0, and elements with atomic number less than those of the
heavy neutron-capture elements, there is a majority fraction that
exhibits well-defined trends with small dispersion of [X/Fe], together
with a minority, C-rich, one that becomes more prominent as one
proceeds to lowest abundance and exhibits large enhancements, to
varying degrees, of the other lighter elements (N, O, Na, Mg, and Al).

\subsubsection{Carbon, Nitrogen, and Oxygen}\label{CNO}

Inspection of the CNO data for {\hes} leads to our second important
conclusion.  {\it With [C/Fe]~=~+1.65, [N/Fe] $<$ +1.47, and [O/Fe]
$<$ +3.09, {\hes} belongs to the C-rich group.  Currently all objects
with [Fe/H] $<$ --4.0 share this property.}

The question of relative carbon and oxygen overabundance at lowest
[Fe/H] has been discussed by \citet{Frebeletal:2007c}, who address the
difficulty of forming low-mass stars, such as {\hen} and {\hef}, in
the early Universe.  They argue that cooling by C and O (via
fine-structure lines of \ion{C}{2} and \ion{O}{1}) provides the means
of forming objects with M $<$ 1~M$_{\odot}$.  They predict that the
formation of such low-mass stars will occur due to cooling by C and O
if the ``transition discriminant" D$_{\rm trans}$ satisfies the
condition: \begin{displaymath} D_{\rm trans}\equiv \log_{10}\left(
10^{\scriptsize \mbox{[C/H]}} + 0.3\times 10^{\scriptsize
\mbox{[O/H]}}\right)\mbox{\ $>$ --3.5,} \end{displaymath}

As may be seen from their Figure 1, both {\hen} and {\hef} (with
D$_{\rm trans} \sim$ --2.7 to --2.4, respectively) easily satisfy the
condition.  For {\hes} we have a relative overabundance of carbon but
only a weak constraint on that of oxygen.  If we assume that the
latter lies in the range zero to $\log\epsilon_{\rm
3D}~(\mbox{O})\sim$~5.5\footnote{According to \citet{Colletetal:2006}
$\log\epsilon_{\rm 3D}~(\mbox{O})\sim$~5.0 and 6.0 in {\hen} and
{\hef}, respectively.}, we obtain --4.1 $< D_{\rm trans} <$ --3.5 for
$\teffm~=~4900$\,K. (We adopt $\log\epsilon_{\rm 3D}(\mbox{C}) -
\log\epsilon_{\rm 1D}(\mbox{C}) = -1.0$, following
\citet{Colletetal:2006} for {\hen}.)  This range lies close to the
limit, but is not inconsistent with the hypothesis of
\citet{Frebeletal:2007c}.  That said, more work is clearly needed to
further constrain the abundances of O in {\hes}.  A low value for
$\log\epsilon$~(O) could marginally challenge their prediction.

\subsubsection{The Heavier Elements}

Closer inspection of Figure~\ref{Fig:Relative_Abundances} leads to a
less exotic, but nevertheless important conclusion. {\it For all
elements other than C, with the possible exception of Co, {\hes} has
[X/Fe] values that differ little from those found in the majority
fraction described above.}  In this regard {\hes} is similar to
HE~1300+0157 ([Fe/H] = --3.88, \citealt{Frebeletal:2007a}),
represented in the figure by the open star.  Given the combination of
carbon enhancement and low abundances of the heavy neutron-capture
elements, it is evident that {\hes} belongs to the CEMP (Carbon
Enhanced Metal Poor)-no group \citep[see][]{Beers/Christlieb:2005},
which have carbon enhancement and [Ba/Fe]~$<$~0.0 (as does
HE~1300+0157).

\subsubsubsection{The Iron Peak}\label{ironpeak}

We draw the reader's attention to the apparently low value of [Co/Fe]
for {\hes} with respect to the bulk of the comparison objects in
Figure~\ref{Fig:Relative_Abundances}.  While one sees, for Cr/Fe and
Mn/Fe, that {\hes} displays the relative underabundances, and for
Ni/Fe the solar value, consistent with the trends for extremely
metal-poor stars first reported by \citet{McWilliametal:1995b} and
\citet{Ryanetal:1996}, Co/Fe appears to behave differently\footnote{We
note here for completeness that our spectra, unfortunately, do not
encompass the important \ion{Zn}{1} lines at 4722.15 and
4810.53\,{\AA}, and we are thus unable to seek to determine or set
limits on the abundance of this important element.}.  That is to say,
for {\hes} one sees [Co/Fe] = 0.0, somewhat lower than the value
[Co/Fe] $\sim$~0.5 that pertains to the bulk of objects with [Fe/H]
$<$~--3.5.  A word of caution is, however, perhaps in order:
inspection of Table~\ref{Tab:1DLTEAbundances} shows that for
$\teffm~=~5100$\,K one obtains [Co/Fe] = 0.3 for {\hes}.  The reader
should bear this in mind in the following discussion.

Possible explanations of the relative underabundances of Cr and Mn,
the relative overabundance of Co, and the solar behaviour of Ni/Fe for
the bulk of extremely metal-poor stars (which demonstrate the need for
a possibly wide range of conditions during the explosions of the first
stars) include higher than canonical explosion energies (i.e.,
``hypernovae", with $>$~10$^{51}$ ergs) and smaller mass cuts than
usually adopted to produce sufficient Fe.  A third relevant effect is
variation in Y$_{\it e}$, the electron mole number\footnote{Y$_{\it
e}$ = $\Sigma$Z$_{j}$X$_{j}$/A$_{j}$, where Z$_{j}$, X$_{j}$, and
A$_{j}$ are the atomic number, mass fraction, and mass number of each
species, respectively.  Neutron excess is then defined as $\eta$ = 1
-- 2Y$_{\it e}$}.  We refer the reader to \citet{Umeda/Nomoto:2002,
Umeda/Nomoto:2005} and \citet{Nomotoetal:2006} for a comprehensive
discussion of these points.  Suffice it here to say that variation of
the explosion energy and of the mass cut can each act to produce the
observed coupled relative underabundance behaviour of [Cr/Fe] and
[Mn/Fe], on the one hand, and the relative overabundance one of
[Co/Fe], on the other.  It is difficult, then, to see how the
departure of the [Co/Fe] value in {\hes} from the general trend might
be explained in this manner.  In some contrast, varying Y$_{\it e}$ on
the range 0.4995--0.5005 in an object with M = 25~M$_\odot$,
Z~=~10$^{-4}$, and E = 2$\times$10$^{52}$ ergs leads to strong
decoupling of the nexus between these elements \citep[][ e.g., Figure
4]{Umeda/Nomoto:2005}. We shall return to [Co/Fe] (and [Cr/Fe]) in
{\hes} in \S\ref{popii}.

\subsection{On the Origin of {\hes}} 

Given the similarity between the relative abundances of {\hes} and
HE~1300+0157 (their iron abundance difference $\Delta$[Fe/H] = 0.9
notwithstanding) we refer the reader to the comprehensive work of
\citet{Frebeletal:2007a} who have considered the various possible
origins of the latter, and which has great relevance to the present
discussion.

\subsubsection {The Role of Binarity}

The interpretation of the abundance patterns of the most metal-poor
stars turns to some extent on their multiplicity.  For example, while
no significant radial velocity variations have been observed for
either of the two most metal-poor stars {\hen} and {\hef} ([Fe/H]
$\sim$ --5.3), in the case of the former, \citet{Sudaetal:2004} have
proposed a Population III, binary model in which {\hen} is a first
generation zero heavy-element object that has experienced binary mass
transfer from an asymptotic giant branch (AGB) star, together with
accretion from the interstellar medium.  A Population II variation on
this theme is that {\hes} could be a second- or later- generation
object that has experienced mass transfer from an erstwhile more
massive (possibly AGB) companion, leading to enrichment of carbon but
not of the heavy neutron-capture elements. See \citet{Komiyaetal:2007}
for a discussion of this mechanism.

In neither case can such models be excluded as an explanation of
{\hes} by the current level of non-detection of radial velocity
variations -- as reported in \S\ref{radial_velocities}, none has been
detected in the present, albeit sparse, data. Future continued radial
velocity monitoring is needed to potentially reach a definitive
conclusion on its multiplicity.

\subsubsection {{\hes} as a Second or Later Generation Object}\label{popii}

The interpretation of {\hes} with [Fe/H]~=~--4.8, and [C/Fe]~=~+1.6 as
a single, Population II, object requires a scenario in which carbon
has been overproduced relative to the heavier elements.
 
While for {\hes} one seeks to understand the large C enhancement, our
current lack of knowledge concerning the abundances of N and O
suggests that one might also consider a more general solution
(applicable not only to C but also other light elements) that involves
the preferential expulsion, during the explosion of objects with
masses greater than $\sim$~10 M$_\odot$, of the outer layers (in which
the lighter elements but not the heavier ones have been synthesized),
but in which only a small fraction of the inner regions (where the
heavier elements have also been produced) is expelled.  Some
suggestions/models that offer promise in this regard are:

$\bullet$ Rotating, zero-heavy-element (Z~=~0), very massive (250--300
M$_\odot$) objects \citep{Fryeretal:2001}

$\bullet$ Failed (low energy), Z = 0, supernovae (see e.g.
\citealt{Woosley/Weaver:1995} and \citealt{Limongietal:2003})

$\bullet$ Z~=~0, 20--30 M$_\odot$, ``faint" supernovae (explosion
energies lower than 10$^{51}$ ergs) with ``mixing and fallback" (see
\citealt{Umeda/Nomoto:2003} and \citealt{Iwamotoetal:2005})

$\bullet$ Rotating, Z~=~0, 60 M$_\odot$ stars \citep{Meynetetal:2006}

In the context of objects forming from the ejecta of such putative
objects, one has also to consider its possible admixture with the
existing interstellar medium, which may have already been enriched by
the ejecta of less exotic objects, among which one might include
non-rotating, energetic core-collapse and pair-instability supernovae
in the mass range 10 $<$ M/M$_\odot$ $<$300 (see
e.g. \citealt{Woosley/Weaver:1995}; \citealt{Umeda/Nomoto:2002};
\citealt{Heger/Woosley:2002}).

Given the approximate nature of the above one dimensional models, it
remains to be seen which, if any, of them might be relevant to the
enrichment of the material from which {\hes} and the other C-rich most
metal-poor stars formed.  That said, the ``mixing and fallback" models
of \citet{Umeda/Nomoto:2003,Umeda/Nomoto:2005} reproduce most of the
abundance patterns seen in the objects at [Fe/H] $\sim$ --5.3.  The
growing number of objects with [Fe/H] $<$ --3.5 that exhibit relative
overabundances of the light elements, together with the diversity of
their abundance patterns, are now establishing a challenging framework
against which to test and constrain such putative theoretical models.

In the context of {\hes}, stronger limits on oxygen and nitrogen are
now required to more rigorously constrain our understanding of its
formation and evolution. That said, we noted above that
\citet{Frebeletal:2007a} have offered a comprehensive critique of the
various models relevant to the manner in which HE~1300+0157 may have
formed, which is useful in informing the present discussion of {\hes}.
Their basic conclusion was that the CEMP-no abundance pattern of
HE~1300+0157 could perhaps result in either of two ways.

Their first scenario envisages enrichment from a single ``faint"
supernova, with ``mixing and fallback" (explosion energies lower than
10$^{51}$ ergs, see e.g. \citealt{Umeda/Nomoto:2003} and
\citealt{Iwamotoetal:2005}), ``with an unusual mass cut allowing for
excesses in C and O".  It will be interesting to see whether this
versatile type of model can explain the Fe-peak data for {\hes} in
Figure~\ref{Fig:Relative_Abundances}.  Certainly the mixing and
fallback model of \citet{Umeda/Nomoto:2003} for the giant {\hen} (with
similar $\teffm$ and log$g$ to those of {\hes}), with [Co/Fe] $\sim$
--1.6, more than overcomes the problem of low [Co/Fe] for {\hes}
discussed in \S\ref{ironpeak}. On the other hand, in that model one
finds the prediction of [Cr/Fe] $\sim$~0.3, considerably higher than
[Cr/Fe]$_{\rm 3D}$ = --1.1 reported here for {\hes}\footnote{From
columns (6) and (8) of Table~\ref{Tab:1DLTEAbundances}.}.

Their second suggestion postulates two enrichment sources, with the
admixture of ejecta from a massive slowly rotating Population III
object (M $>$ 100 M$_\odot$, Z = 0; see
e.g. \citealt{Heger/Woosley:2002}) which produces copious amounts of C
and O, but little N and heavier elements before exploding as a
pair-instability SN, together with material from a ``hypernova"
(explosion energy E $\sim$ 10$\times$10$^{51}$ ergs, see e.g.
\citealt{Umeda/Nomoto:2005}, and \citealt{Nomotoetal:2006}) which
would contribute heavier elements having atomic number less than 30.
There are, however, problems with this model as an explanation of
{\hes}.  While it is consistent with the observation of [Co/Fe]~=~0.76
in HE~1300+0157, it faces difficulty as an explanation of
[Co/Fe]~=~0.0 ([Co/Fe]$_{\rm 3D}$~=~--0.3) in {\hes}, since, as
discussed above in \S\ref{ironpeak}, the hypernovae of
\citet{Umeda/Nomoto:2005} and \citet{Nomotoetal:2006} produce higher
Co abundances, for which they report [Co/Fe]~$\sim$~0.4.

\section{Summary}\label{summary}

We have reported the discovery and analysis of the ultra metal-poor
red giant\\ {\hes}, which has atmospheric parameters
{\tefft}~=~4900\,K, $\log g$~=~2.2, and [Fe/H] = --4.75.  This object
is also carbon rich, with [C/Fe]~=~+1.6, a property shared by the
three objects currently known to have iron abundance less than [Fe/H]
$\sim$ --4.0.  We have also obtained relative abundances, or abundance
limits, for a further 18 elements, and find that, with the possible
exception of [Co/Fe], {\hes} has an abundance pattern similar to that
of the majority of objects with [Fe/H] $\sim$ --4.0.

{\hes} falls in the ``gap'' between the two most metal-poor stars at
[Fe/H] $\sim$ --5.5 and the next most metal-poor stars at [Fe/H ] =
--4.0.  Its discovery weakens the case for the existence of such a
``gap".  (That said, while the statistics are poor, the number of
objects with --5.0 $<$ [Fe/H] $<$ --4.0 is lower by a factor
$\sim$3--4 than predicted at the earliest times by existing models of
galactic chemical enrichment.)

The abundance pattern of {\hes} has been discussed in the context of
single and two-component enrichment models. Existing ``mixing and
fallback'' models provide a poor explanation of the observed relative
abundances, while a two-component model provides a better, but still
inadequate one.

The overabundance of carbon in {\hes} is (marginally) consistent with
the hypothesis of \citet{Frebeletal:2007c} that carbon (and oxygen)
richness may be a necessary requirement for formation of low mass
stars at the earliest times.

Further data are required to determine the oxygen abundance and
improve that of Co, and hence more strongly constrain the origin of
{\hes}.

\acknowledgements
 
We would like to thank the Director's Discretionary Time Committee of
ESO for allocating time to this project. We express our gratitude to
the ESO staff on Paranal and in Garching for carrying out the
observations, and reducing the data, respectively. We are grateful to
Thomas Szeifert for making it possible for the observations of {\hen}
to be initiated on the night of 5 February, 2006, and to B. Plez and
A. Jorissen for providing us with $^{12}$CH and $^{13}$CH line
lists. We also thank Brian Schmidt for beneficial discussion on the
kinematics of {\hes}.

J.E.N and M.S.B acknowledge support from the Australian Research
Council under grants DP0342613 and DP0663562. N.C. is a Research
Fellow of the Royal Swedish Academy of Sciences supported by a grant
from the Knut and Alice Wallenberg Foundation. He also acknowledges
financial support from Deutsche Forschungsgemeinschaft through grants
Ch~214/3 and Re~353/44. A.J.K. and K.E. acknowledge support from
Swedish Research Council grants 50349801 and 60307001, respectively.
T.C.B. acknowledges partial funding for this work from grants AST
04-06784 and PHY 02-16873: Physics Frontier Center/ Joint Institute
for Nuclear Astrophysics (JINA), both awarded by the National Science
Foundation.\\\\ .

\noindent{\it Facilities:} {ANU:SSO2.3m(DBS); VLT:Kueyen(UVES)}


\begin{thebibliography}{75}
\expandafter\ifx\csname natexlab\endcsname\relax\def\natexlab#1{#1}\fi

\bibitem[{Alonso {et~al.}(1998)Alonso, Arribas, \&
  {Mart{\'\i}nez-Roger}}]{Alonsoetal:1998}
Alonso, A., Arribas, S., \& {Mart{\'\i}nez-Roger}, C. 1998, A\&AS, 131, 209

\bibitem[{Alonso {et~al.}(1999)Alonso, Arribas, \&
  {Mart{\'\i}nez-Roger}}]{Alonsoetal:1999b}
---. 1999, A\&AS, 140, 261

\bibitem[{Alonso {et~al.}(2001)Alonso, Arribas, \&
  Mart{\'\i}nez-Roger}]{Alonsoetal:2001}
Alonso, A., Arribas, S., \& Mart{\'\i}nez-Roger, C. 2001, A\&A, 376, 1039

\bibitem[{Aoki {et~al.}(2006)Aoki, Frebel, Christlieb, Norris, Beers, Minezaki,
  Barklem, Honda, Takada-Hidai, Asplund, Ryan, Tsangarides, Eriksson,
  Steinhauer, Deliyannis, Nomoto, Fujimoto, Ando, Yoshii, \&
  Kajino}]{Aokietal:2006}
Aoki, W., Frebel, A., Christlieb, N., Norris, J.~E., Beers, T.~C., Minezaki,
  T., Barklem, P.~S., Honda, S., Takada-Hidai, M., Asplund, M., Ryan, S.~G.,
  Tsangarides, S., Eriksson, K., Steinhauer, A., Deliyannis, C.~P., Nomoto, K.,
  Fujimoto, M.~Y., Ando, H., Yoshii, Y., \& Kajino, T. 2006, ApJ, 639, 897

\bibitem[{Aoki {et~al.}(2002)Aoki, Norris, Ryan, Beers, \&
  Ando}]{Aokietal:2002d}
Aoki, W., Norris, J.~E., Ryan, S.~G., Beers, T.~C., \& Ando, H. 2002, ApJ, 576,
  L141

\bibitem[{Aoki {et~al.}(2004)Aoki, Norris, Ryan, Beers, Christlieb,
  Tsangarides, \& Ando}]{Aokietal:2004a}
Aoki, W., Norris, J.~E., Ryan, S.~G., Beers, T.~C., Christlieb, N.,
  Tsangarides, S., \& Ando, H. 2004, ApJ, 608, 971

\bibitem[{Asplund {et~al.}(2005)Asplund, Grevesse, \&
  Sauval}]{Asplundetal:2005a}
Asplund, M., Grevesse, N., \& Sauval, A.~J. 2005, in ASP Conf. Ser., Vol. 336,
  Cosmic Abundances as Records of Stellar Evolution and Nucleosynthesis in
  honor of David L. Lambert, ed. T.~G. Barnes \& F.~N. Bash, San Francisco, 25

\bibitem[{Barklem {et~al.}(2000)Barklem, Piskunov, \&
  O'Mara}]{Barklemetal:2000a}
Barklem, P.~S., Piskunov, N., \& O'Mara, B.~J. 2000, A\&A, 363, 1091

\bibitem[{Beers \& Christlieb(2005)}]{Beers/Christlieb:2005}
Beers, T.~C. \& Christlieb, N. 2005, ARA\&A, 43, 531

\bibitem[{Beers {et~al.}(2005)Beers, Christlieb, Norris, Bessell, Wilhelm,
  {Allende Prieto}, Yanny, Rockosi, Newberg, Rossi, \& Lee}]{Beersetal:2005a}
Beers, T.~C., Christlieb, N., Norris, J.~E., Bessell, M.~S., Wilhelm, R.,
  {Allende Prieto}, C., Yanny, B., Rockosi, C., Newberg, H.~J., Rossi, S., \&
  Lee, Y.~S. 2005, in IAU Symposium, Vol. 228, From Lithium to Uranium:
  Elemental Tracers of Early Cosmic Evolution, ed. V.~Hill, P.~Fran{\c c}ois,
  \& F.~Primas, 175

\bibitem[{Beers {et~al.}(2007)Beers, Flynn, Rossi, Sommer-Larsen, Wilhelm,
  Marsteller, Lee, {De Lee}, Krugler, Deliyannis, Simmons, Mills, Zickgraf,
  Holmberg, {\"O}nehag, Eriksson, Terndrup, Salim, Andersen, Nordstr{\"o}m,
  Christlieb, Frebel, \& Rhee}]{Beersetal:2007}
Beers, T.~C., Flynn, C., Rossi, S., Sommer-Larsen, J., Wilhelm, R., Marsteller,
  B., Lee, Y.~S., {De Lee}, N., Krugler, J., Deliyannis, C.~P., Simmons, A.~T.,
  Mills, E., Zickgraf, F.-J., Holmberg, J., {\"O}nehag, A., Eriksson, A.,
  Terndrup, D.~M., Salim, S., Andersen, J., Nordstr{\"o}m, B., Christlieb, N.,
  Frebel, A., \& Rhee, J. 2007, ApJ Suppl., 168, 128

\bibitem[{Bessell {et~al.}(1998)Bessell, Castelli, \& Plez}]{Besselletal:1998}
Bessell, M.~S., Castelli, F., \& Plez, B. 1998, A\&, 333, 231

\bibitem[{Bessell {et~al.}(2004)Bessell, Christlieb, \&
  Gustafsson}]{Besselletal:2004}
Bessell, M.~S., Christlieb, N., \& Gustafsson, B. 2004, ApJ, 612, L61

\bibitem[{Carpenter(2001)}]{Carpenter:2001}
Carpenter, J.~M. 2001, AJ, 121, 2851

\bibitem[{Carretta {et~al.}(2002)Carretta, Gratton, Cohen, Beers, \&
  Christlieb}]{KeckpaperII}
Carretta, E., Gratton, R., Cohen, J.~G., Beers, T.~C., \& Christlieb, N. 2002,
  AJ, 124, 481

\bibitem[{Cayrel {et~al.}(2004)Cayrel, Depagne, Spite, Hill, Spite, Fran{\c
  c}ois, Beers, Primas, Andersen, Barbuy, Bonifacio, Molaro, \&
  Nordstr\"om}]{Cayreletal:2004}
Cayrel, R., Depagne, E., Spite, M., Hill, V., Spite, F., Fran{\c c}ois, P.,
  Beers, T.~C., Primas, F., Andersen, J., Barbuy, B., Bonifacio, P., Molaro,
  P., \& Nordstr\"om, B. 2004, A\&A, 416, 1117

\bibitem[{Chamberlain \& Aller(1951)}]{Chamberlain/Aller:1951}
Chamberlain, J.~W. \& Aller, L.~H. 1951, ApJ, 114, 52

\bibitem[{Christlieb {et~al.}(2002)Christlieb, Bessell, Beers, Gustafsson,
  Korn, Barklem, Karlsson, Mizuno-Wiedner, \& Rossi}]{HE0107_Nature}
Christlieb, N., Bessell, M.~S., Beers, T.~C., Gustafsson, B., Korn, A.,
  Barklem, P.~S., Karlsson, T., Mizuno-Wiedner, M., \& Rossi, S. 2002, Nature,
  419, 904

\bibitem[{Christlieb {et~al.}(2004)Christlieb, Gustafsson, Korn, Barklem,
  Beers, Bessell, Karlsson, \& Mizuno-Wiedner}]{HE0107_ApJ}
Christlieb, N., Gustafsson, B., Korn, A.~J., Barklem, P.~S., Beers, T.~C.,
  Bessell, M.~S., Karlsson, T., \& Mizuno-Wiedner, M. 2004, ApJ, 603, 708

\bibitem[{{Christlieb~et~al.}(2007)}]{Christliebetal:2007}
{Christlieb,~N.~et~al.} 2007, in preparation

\bibitem[{Clarke \& Bromm(2003)}]{Clarke/Bromm:2003}
Clarke, C.~J. \& Bromm, V. 2003, MNRAS, 343, 1224

\bibitem[{Cohen {et~al.}(2004)Cohen, Christlieb, McWilliam, Shectman, Thompson,
  Wasserburg, Ivans, Dehn, Karlsson, \& Mel{\' e}ndez}]{KeckpaperIV}
Cohen, J.~G., Christlieb, N., McWilliam, A., Shectman, S., Thompson, I.,
  Wasserburg, G.~J., Ivans, I.~I., Dehn, M., Karlsson, T., \& Mel{\' e}ndez, J.
  2004, ApJ, 612, 1107

\bibitem[{Collet {et~al.}(2006)Collet, Asplund, \&
  Trampedach}]{Colletetal:2006}
Collet, R., Asplund, M., \& Trampedach, R. 2006, ApJ, 644, L121

\bibitem[{Fran{\c c}ois {et~al.}(2003)Fran{\c c}ois, Depagne, Hill, Spite,
  Spite, Plez, Beers, Barbuy, Cayrel, Andersen, Bonifacio, Molaro, Nordstr\"om,
  \& Primas}]{Francoisetal:2003}
Fran{\c c}ois, P., Depagne, E., Hill, V., Spite, M., Spite, F., Plez, B.,
  Beers, T.~C., Barbuy, B., Cayrel, R., Andersen, J., Bonifacio, P., Molaro,
  P., Nordstr\"om, B., \& Primas, F. 2003, A\&A, 403, 1105

\bibitem[{Frebel {et~al.}(2005)Frebel, Aoki, Christlieb, Ando, Asplund,
  Barklem, Beers, Eriksson, Fechner, Fujimoto, Honda, Kajino, Minezaki, Nomoto,
  Norris, Ryan, Takada-Hidai, Tsangarides, \& Yoshii}]{Frebeletal:2005}
Frebel, A., Aoki, W., Christlieb, N., Ando, H., Asplund, M., Barklem, P.~S.,
  Beers, T.~C., Eriksson, K., Fechner, C., Fujimoto, M.~Y., Honda, S., Kajino,
  T., Minezaki, T., Nomoto, K., Norris, J.~E., Ryan, S.~G., Takada-Hidai, M.,
  Tsangarides, S., \& Yoshii, Y. 2005, Nature, 434, 871

\bibitem[{Frebel {et~al.}(2006)Frebel, Christlieb, Norris, Aoki, \&
  Asplund}]{Frebeletal:2006}
Frebel, A., Christlieb, N., Norris, J.~E., Aoki, W., \& Asplund, M. 2006, ApJL,
  638, L17

\bibitem[{Frebel {et~al.}(2007{\natexlab{a}})Frebel, Johnson, \&
  Bromm}]{Frebeletal:2007c}
Frebel, A., Johnson, J.~L., \& Bromm, V. 2007{\natexlab{a}}, ApJ, submitted,
  astro-ph/0701139

\bibitem[{Frebel {et~al.}(2007{\natexlab{b}})Frebel, Norris, Aoki, Honda,
  Bessell, Takada-Hidai, Beers, \& Christlieb}]{Frebeletal:2007a}
Frebel, A., Norris, J.~E., Aoki, W., Honda, S., Bessell, M.~S., Takada-Hidai,
  M., Beers, T.~C., \& Christlieb, N. 2007{\natexlab{b}}, ApJ, in press,
  astro-ph/0612160

\bibitem[{Fryer {et~al.}(2001)Fryer, Woosley, \& Heger}]{Fryeretal:2001}
Fryer, C.~L., Woosley, S.~E., \& Heger, A. 2001, ApJ, 550, 372

\bibitem[{Fuhrmann {et~al.}(1993)Fuhrmann, Axer, \& Gehren}]{Fuhrmannetal:1993}
Fuhrmann, K., Axer, M., \& Gehren, T. 1993, A\&A, 271, 451

\bibitem[{Gustafsson {et~al.}(2007)Gustafsson, Edvardsson, Eriksson,
  J{\o}rgensen, Nordlund, \& Plez}]{Gustafssonetal:2007}
Gustafsson, B., Edvardsson, B., Eriksson, K., J{\o}rgensen, U.~G., Nordlund,
  {\AA}., \& Plez, B. 2007, in preparation

\bibitem[{Hartwick(1976)}]{Hartwick:1976}
Hartwick, F.~D.~A. 1976, ApJ, 209, 418

\bibitem[{Heger \& Woosley(2002)}]{Heger/Woosley:2002}
Heger, A. \& Woosley, S.~E. 2002, ApJ, 567, 532

\bibitem[{Hill {et~al.}(2002)Hill, Plez, Cayrel, Beers, Nordstr{\" o}m,
  Andersen, Spite, Spite, Barbuy, Bonifacio, Depagne, Fran{\c c}ois, \&
  Primas}]{Hilletal:2002}
Hill, V., Plez, B., Cayrel, R., Beers, T.~C., Nordstr{\" o}m, B., Andersen, J.,
  Spite, M., Spite, F., Barbuy, B., Bonifacio, P., Depagne, E., Fran{\c c}ois,
  P., \& Primas, F. 2002, A\&A, 387, 560

\bibitem[{Houdashelt {et~al.}(2000)Houdashelt, Bell, \&
  Sweigart}]{Houdasheltetal:2000}
Houdashelt, M.~L., Bell, R.~A., \& Sweigart, A.~V. 2000, AJ, 119, 1448

\bibitem[{Iben(1983)}]{Iben:1983}
Iben, I.~Jr. 1983, Memorie della Societa Astronomica Italiana, 54, 321

\bibitem[{Iwamoto {et~al.}(2005)Iwamoto, Umeda, Tominaga, Nomoto, \&
  Maeda}]{Iwamotoetal:2005}
Iwamoto, N., Umeda, H., Tominaga, N., Nomoto, K., \& Maeda, K. 2005, Science,
  309, 451

\bibitem[{Johnson {et~al.}(2007)Johnson, Herwig, Beers, \&
  Christlieb}]{Johnsonetal:2007}
Johnson, J.~J., Herwig, F., Beers, T.~C., \& Christlieb, N. 2007, ApJ, in
  press, astro-ph/0608666

\bibitem[{J{\o}rgensen {et~al.}(1996)J{\o}rgensen, Larsson, Iwamae, \&
  Yu}]{Jorgensenetal:1996}
J{\o}rgensen, U.~G., Larsson, M., Iwamae, A., \& Yu, B. 1996, A\&A, 315, 204

\bibitem[{Karlsson(2006)}]{Karlsson:2006}
Karlsson, T. 2006, ApJ, 641, L41

\bibitem[{Komiya {et~al.}(2007)Komiya, Suda, Minaguchi, Shigeyama, Aoki, \&
  Fujimoto}]{Komiyaetal:2007}
Komiya, Y., Suda, T., Minaguchi, H., Shigeyama, T., Aoki, W., \& Fujimoto, M.
  2007, ApJ, in press, astro-ph/0610670

\bibitem[{Korn {et~al.}(2006)Korn, Grundahl, Richard, Barklem, Mashonkina,
  Collet, Piskunov, \& Gustafsson}]{Kornetal:2006}
Korn, A.~J., Grundahl, F., Richard, O., Barklem, P.~S., Mashonkina, L., Collet,
  R., Piskunov, N., \& Gustafsson, B. 2006, Nature, 442, 657

\bibitem[{Korn {et~al.}(2003)Korn, Shi, \& Gehren}]{Kornetal:2003}
Korn, A.~J., Shi, J., \& Gehren, T. 2003, A\&A, 407, 691

\bibitem[{Kupka {et~al.}(1999)Kupka, Piskunov, Ryabchikova, Stempels, \&
  Weiss}]{VALD2a}
Kupka, F.~G., Piskunov, N., Ryabchikova, T.~A., Stempels, H.~C., \& Weiss,
  W.~W. 1999, A\&AS, 138, 119

\bibitem[{Kupka {et~al.}(2000)Kupka, Ryabchikova, Piskunov, Stempels, \&
  Weiss}]{VALD2b}
Kupka, F.~G., Ryabchikova, T.~A., Piskunov, N., Stempels, H.~C., \& Weiss,
  W.~W. 2000, Baltic Astronomy, 9, 590

\bibitem[{Kurucz(2006)}]{Kurucz:2006}
Kurucz, R. 2006, http://kurucz.harvard.edu/LINELISTS/LINESMOL

\bibitem[{Limongi {et~al.}(2003)Limongi, Chieffi, \&
  Bonifacio}]{Limongietal:2003}
Limongi, M., Chieffi, A., \& Bonifacio, P. 2003, ApJ, 594, L123

\bibitem[{Luque \& Crosley(1999)}]{Luque/Crosley:1999}
Luque, J. \& Crosley, D.~R. 1999, LIFBASE version 1.5, Sri international report
  mp 99-009, SRI International, http://www.sri.com/cem/lifbase/Lifbase.PDF

\bibitem[{Mashonkina {et~al.}(2007)Mashonkina, Korn, \&
  Przybilla}]{Mashonkinaetal:2007}
Mashonkina, L., Korn, A.~J., \& Przybilla, N. 2007, A\&A, 461, 261

\bibitem[{McWilliam {et~al.}(1995)McWilliam, Preston, Sneden, \&
  Searle}]{McWilliametal:1995b}
McWilliam, A., Preston, G.~W., Sneden, C., \& Searle, L. 1995, AJ, 109, 2757

\bibitem[{Meynet {et~al.}(2006)Meynet, Ekstr{\"o}m, \&
  Maeder}]{Meynetetal:2006}
Meynet, G., Ekstr{\"o}m, S., \& Maeder, A. 2006, A\&A, 447, 623

\bibitem[{{Napiwotzki~et~al.}(2007)}]{Napiwotzkietal:2007}
{Napiwotzki,~R.~et~al.} 2007, in preparation

\bibitem[{Nomoto {et~al.}(2006)Nomoto, Tominaga, Umeda, Kobayashi, \&
  Maeda}]{Nomotoetal:2006}
Nomoto, K., Tominaga, N., Umeda, H., Kobayashi, C., \& Maeda, K. 2006,
  Nuc.Phys.A, 777, 424, astro-ph/0605725

\bibitem[{Norris {et~al.}(2000)Norris, Beers, \& Ryan}]{Norrisetal:2000}
Norris, J.~E., Beers, T.~C., \& Ryan, S.~G. 2000, ApJ, 540, 456

\bibitem[{Norris {et~al.}(1997)Norris, Ryan, \& Beers}]{Norrisetal:1997b}
Norris, J.~E., Ryan, S.~G., \& Beers, T.~C. 1997, ApJ, 489, L169

\bibitem[{Norris {et~al.}(2001)Norris, Ryan, \& Beers}]{Norrisetal:2001}
---. 2001, ApJ, 561, 1034

\bibitem[{Norris {et~al.}(2002)Norris, Ryan, Beers, Aoki, \&
  Ando}]{Norrisetal:2002}
Norris, J.~E., Ryan, S.~G., Beers, T.~C., Aoki, W., \& Ando, H. 2002, ApJ, 569,
  L107

\bibitem[{Plez \& Cohen(2005)}]{Plez/Cohen:2005}
Plez, B. \& Cohen, J.~G. 2005, A\&A, 434, 1117

\bibitem[{Prantzos(2003)}]{Prantzos:2003}
Prantzos, N. 2003, A\&A, 404, 211

\bibitem[{Ryan \& Norris(1991)}]{Ryan/Norris:1991}
Ryan, S.~G. \& Norris, J.~E. 1991, AJ, 101, 1865

\bibitem[{Ryan {et~al.}(1996)Ryan, Norris, \& Beers}]{Ryanetal:1996}
Ryan, S.~G., Norris, J.~E., \& Beers, T.~C. 1996, ApJ, 471, 254

\bibitem[{Ryan {et~al.}(1991)Ryan, Norris, \& Bessell}]{Ryanetal:1991}
Ryan, S.~G., Norris, J.~E., \& Bessell, M.~S. 1991, AJ, 102, 303

\bibitem[{Salvadori {et~al.}(2007)Salvadori, Schneider, \&
  Ferrara}]{Salvadorietal:2007}
Salvadori, S., Schneider, R., \& Ferrara, A. 2007, MNRAS, submitted,
  astro-ph/0611130

\bibitem[{Schlegel {et~al.}(1998)Schlegel, Finkbeiner, \&
  Davis}]{Schlegeletal:1998}
Schlegel, D.~J., Finkbeiner, D.~P., \& Davis, M. 1998, ApJ, 500, 525

\bibitem[{Searle \& Sargent(1972)}]{Searle/Sargent:1972}
Searle, L. \& Sargent, W.~L.~W. 1972, ApJ, 173, 25

\bibitem[{{Seaton} {et~al.}(1994){Seaton}, {Yan}, {Mihalas}, \&
  {Pradhan}}]{Seatonetal:1994}
{Seaton}, M.~J., {Yan}, Y., {Mihalas}, D., \& {Pradhan}, A.~K. 1994, MNRAS,
  266, 805

\bibitem[{{Skrutskie} {et~al.}(2006){Skrutskie}, {Cutri}, {Stiening},
  {Weinberg}, {Schneider}, {Carpenter}, {Beichman}, {Capps}, {Chester},
  {Elias}, {Huchra}, {Liebert}, {Lonsdale}, {Monet}, {Price}, {Seitzer},
  {Jarrett}, {Kirkpatrick}, {Gizis}, {Howard}, {Evans}, {Fowler}, {Fullmer},
  {Hurt}, {Light}, {Kopan}, {Marsh}, {McCallon}, {Tam}, {Van Dyk}, \&
  {Wheelock}}]{Skrutskieetal:2006}
{Skrutskie}, M.~F., {Cutri}, R.~M., {Stiening}, R., {Weinberg}, M.~D.,
  {Schneider}, S., {Carpenter}, J.~M., {Beichman}, C., {Capps}, R., {Chester},
  T., {Elias}, J., {Huchra}, J., {Liebert}, J., {Lonsdale}, C., {Monet}, D.~G.,
  {Price}, S., {Seitzer}, P., {Jarrett}, T., {Kirkpatrick}, J.~D., {Gizis},
  J.~E., {Howard}, E., {Evans}, T., {Fowler}, J., {Fullmer}, L., {Hurt}, R.,
  {Light}, R., {Kopan}, E.~L., {Marsh}, K.~A., {McCallon}, H.~L., {Tam}, R.,
  {Van Dyk}, S., \& {Wheelock}, S. 2006, AJ, 131, 1163

\bibitem[{{Smith} \& {Gallagher}(1966)}]{Smith/Gallagher:1966}
{Smith}, W.~W. \& {Gallagher}, A. 1966, Phys. Rev., 145, 26

\bibitem[{Suda {et~al.}(2004)Suda, Aikawa, Machida, Fujimoto, \&
  Iben}]{Sudaetal:2004}
Suda, T., Aikawa, M., Machida, M.~N., Fujimoto, M.~Y., \& Iben, I.~Jr. 2004, ApJ,
  611, 476

\bibitem[{Umeda \& Nomoto(2002)}]{Umeda/Nomoto:2002}
Umeda, H. \& Nomoto, K. 2002, ApJ, 565, 385

\bibitem[{Umeda \& Nomoto(2003)}]{Umeda/Nomoto:2003}
---. 2003, Nature, 422, 871

\bibitem[{Umeda \& Nomoto(2005)}]{Umeda/Nomoto:2005}
---. 2005, ApJ, 619, 427

\bibitem[{Wisotzki {et~al.}(2000)Wisotzki, Christlieb, Bade, Beckmann,
  K\"ohler, Vanelle, \& Reimers}]{hespaperIII}
Wisotzki, L., Christlieb, N., Bade, N., Beckmann, V., K\"ohler, T., Vanelle,
  C., \& Reimers, D. 2000, A\&A, 358, 77

\bibitem[{Woosley \& Weaver(1995)}]{Woosley/Weaver:1995}
Woosley, S.~E. \& Weaver, T.~A. 1995, ApJS, 101, 181

\bibitem[{Yoshii(1981)}]{Yoshii:1981}
Yoshii, Y. 1981, A\&A, 97, 280

\end{thebibliography}

\clearpage

\begin{deluxetable}{lllrrrr} 
\tablecolumns{7} 
\tablewidth{0pt} 
\tablecaption{\label{Tab:Linelist} AVERAGED EQUIVALENT WIDTHS, 
  UPPER LIMITS, AND LINE-BY-LINE ABUNDANCES OF {\hes}.}
\tablehead{
  \colhead{}    & \colhead{$\lambda$} & \colhead{$\chi$} & \colhead{$\log gf$} &
  \colhead{$W_{\lambda}$} & \colhead{$\log\epsilon_{\rm 4900\,K}$} & 
  \colhead{$\log\epsilon_{\rm 5100\,K}$}\\
  \colhead{Species} & \colhead{({\AA})}    & \colhead{(eV)}  & \colhead{(dex)}    &
  \colhead{(m{\AA})} & \colhead{(dex)} & \colhead{(dex)}\\
  \colhead{(1)} & \colhead{(2)}    & \colhead{(3)}  & \colhead{(4)}    &
  \colhead{(5)} & \colhead{(6)} & \colhead{(7)}\\
  }
\startdata
\ion{Li}{1} & $6707.761$ & $0.00$ & $-0.009$ & synth & $< 0.70$ & $< 0.90$ \\
\ion{O }{1} & $6300.304$ & $0.00$ & $-9.717$ & $< 5$ & $< 7.00$ & $< 7.13$ \\
\ion{Na}{1} & $5889.951$ & $0.00$ & $ 0.117$ & $ 15$ & $  1.26$ & $  1.44$ \\
\ion{Mg}{1} & $3829.355$ & $2.71$ & $-0.231$ & $ 49$ & $  3.03$ & $  3.19$ \\
\ion{Mg}{1} & $3832.304$ & $2.71$ & $ 0.121$ & $ 68$ & $  3.00$ & $  3.16$ \\
\ion{Mg}{1} & $3838.292$ & $2.72$ & $ 0.392$ & $ 79$ & $  2.95$ & $  3.11$ \\
\ion{Mg}{1} & $5172.684$ & $2.71$ & $-0.399$ & $ 53$ & $  3.12$ & $  3.30$ \\
\ion{Mg}{1} & $5183.604$ & $2.72$ & $-0.177$ & $ 64$ & $  3.06$ & $  3.24$ \\
\ion{Al}{1} & $3961.520$ & $0.01$ & $-0.323$ & $ 19$ & $  1.01$ & $  1.22$ \\
\ion{Ca}{1} & $4226.728$ & $0.00$ & $ 0.265$ & synth & $  1.81$ & $  2.03$ \\
\ion{Ca}{2} & $3706.024$ & $3.12$ & $-0.480$ & synth & $  2.11$ & $  2.12$ \\
\ion{Ca}{2} & $3933.663$ & $0.00$ & $ 0.105$ & synth & $  2.26$ & $  2.51$ \\
\ion{Sc}{2} & $3572.526$ & $0.02$ & $ 0.267$ & $ 25$ & $ -1.54$ & $ -1.37$ \\
\ion{Sc}{2} & $3576.340$ & $0.01$ & $ 0.007$ & $ 21$ & $ -1.38$ & $ -1.21$ \\
\ion{Sc}{2} & $3580.925$ & $0.00$ & $-0.149$ & $ 20$ & $ -1.26$ & $ -1.09$ \\
\ion{Sc}{2} & $3642.784$ & $0.00$ & $ 0.133$ & $ 26$ & $ -1.39$ & $ -1.21$ \\
\ion{Ti}{2} & $3759.296$ & $0.61$ & $ 0.270$ & $ 68$ & $  0.38$ & $  0.51$ \\
\ion{Ti}{2} & $3761.323$ & $0.57$ & $ 0.170$ & $ 65$ & $  0.37$ & $  0.50$ \\
\ion{Ti}{2} & $3900.551$ & $1.13$ & $-0.280$ & $ 21$ & $  0.53$ & $  0.64$ \\
\ion{Ti}{2} & $3913.468$ & $1.12$ & $-0.410$ & $ 20$ & $  0.63$ & $  0.74$ \\
\ion{Ti}{2} & $4395.033$ & $1.08$ & $-0.510$ & $ 17$ & $  0.55$ & $  0.66$ \\
\ion{Ti}{2} & $4443.794$ & $1.08$ & $-0.700$ & $ 12$ & $  0.55$ & $  0.66$ \\
\ion{Cr}{1} & $3578.684$ & $0.00$ & $ 0.409$ & $ 16$ & $  0.19$ & $  0.46$ \\
\ion{Cr}{1} & $4254.332$ & $0.00$ & $-0.114$ & $  7$ & $  0.18$ & $  0.43$ \\
\ion{Mn}{1} & $4030.753$ & $0.00$ & $-0.470$ & $<10$ & $< 0.01$ & $< 0.29$ \\
\ion{Fe}{1} & $3727.619$ & $0.96$ & $-0.631$ & $ 56$ & $  2.69$ & $  2.94$ \\
\ion{Fe}{1} & $3743.362$ & $0.99$ & $-0.785$ & $ 49$ & $  2.71$ & $  2.96$ \\
\ion{Fe}{1} & $3745.561$ & $0.09$ & $-0.771$ & $ 97$ & $  2.91$ & $  3.21$ \\
\ion{Fe}{1} & $3745.900$ & $0.12$ & $-1.335$ & $ 81$ & $  3.04$ & $  3.32$ \\
\ion{Fe}{1} & $3758.233$ & $0.96$ & $-0.027$ & $ 74$ & $  2.48$ & $  2.73$ \\
\ion{Fe}{1} & $3763.789$ & $0.99$ & $-0.238$ & $ 67$ & $  2.55$ & $  2.80$ \\
\ion{Fe}{1} & $3767.192$ & $1.01$ & $-0.389$ & $ 60$ & $  2.56$ & $  2.81$ \\
\ion{Fe}{1} & $3787.880$ & $1.01$ & $-0.859$ & $ 44$ & $  2.70$ & $  2.95$ \\
\ion{Fe}{1} & $3790.093$ & $0.99$ & $-1.761$ & $ 11$ & $  2.75$ & $  3.01$ \\
\ion{Fe}{1} & $3812.964$ & $0.96$ & $-1.047$ & $ 39$ & $  2.72$ & $  2.97$ \\
\ion{Fe}{1} & $3815.840$ & $1.49$ & $ 0.237$ & $ 61$ & $  2.50$ & $  2.73$ \\
\ion{Fe}{1} & $3820.425$ & $0.86$ & $ 0.119$ & $ 88$ & $  2.57$ & $  2.83$ \\
\ion{Fe}{1} & $3824.444$ & $0.00$ & $-1.362$ & $ 85$ & $  2.98$ & $  3.27$ \\
\ion{Fe}{1} & $3825.881$ & $0.92$ & $-0.037$ & $ 76$ & $  2.45$ & $  2.70$ \\
\ion{Fe}{1} & $3827.823$ & $1.56$ & $ 0.062$ & $ 53$ & $  2.59$ & $  2.81$ \\
\ion{Fe}{1} & $3841.048$ & $1.61$ & $-0.045$ & $ 42$ & $  2.52$ & $  2.74$ \\
\ion{Fe}{1} & $3849.967$ & $1.01$ & $-0.871$ & $ 44$ & $  2.69$ & $  2.93$ \\
\ion{Fe}{1} & $3850.818$ & $0.99$ & $-1.734$ & $ 12$ & $  2.76$ & $  3.02$ \\
\ion{Fe}{1} & $3856.372$ & $0.05$ & $-1.286$ & $ 86$ & $  2.98$ & $  3.27$ \\
\ion{Fe}{1} & $3859.911$ & $0.00$ & $-0.710$ & $104$ & $  2.88$ & $  3.19$ \\
\ion{Fe}{1} & $3865.523$ & $1.01$ & $-0.982$ & $ 39$ & $  2.71$ & $  2.96$ \\
\ion{Fe}{1} & $3878.018$ & $0.96$ & $-0.914$ & $ 45$ & $  2.69$ & $  2.94$ \\
\ion{Fe}{1} & $3878.573$ & $0.09$ & $-1.379$ & $ 80$ & $  2.93$ & $  3.22$ \\
\ion{Fe}{1} & $3895.656$ & $0.11$ & $-1.670$ & $ 69$ & $  2.96$ & $  3.24$ \\
\ion{Fe}{1} & $3899.707$ & $0.09$ & $-1.531$ & $ 76$ & $  2.96$ & $  3.24$ \\
\ion{Fe}{1} & $3902.946$ & $1.56$ & $-0.446$ & $ 31$ & $  2.63$ & $  2.86$ \\
\ion{Fe}{1} & $3906.480$ & $0.11$ & $-2.243$ & $ 43$ & $  2.99$ & $  3.27$ \\
\ion{Fe}{1} & $3920.258$ & $0.12$ & $-1.746$ & $ 64$ & $  2.93$ & $  3.21$ \\
\ion{Fe}{1} & $3922.912$ & $0.05$ & $-1.651$ & $ 76$ & $  3.03$ & $  3.31$ \\
\ion{Fe}{1} & $3927.920$ & $0.11$ & $-1.522$ & $ 78$ & $  3.02$ & $  3.31$ \\
\ion{Fe}{1} & $3930.297$ & $0.09$ & $-1.491$ & $ 86$ & $  3.17$ & $  3.46$ \\
\ion{Fe}{1} & $4005.242$ & $1.56$ & $-0.610$ & $ 26$ & $  2.69$ & $  2.91$ \\
\ion{Fe}{1} & $4045.812$ & $1.49$ & $ 0.280$ & $ 69$ & $  2.56$ & $  2.79$ \\
\ion{Fe}{1} & $4063.594$ & $1.56$ & $ 0.062$ & $ 56$ & $  2.60$ & $  2.83$ \\
\ion{Fe}{1} & $4071.738$ & $1.61$ & $-0.022$ & $ 50$ & $  2.61$ & $  2.84$ \\
\ion{Fe}{1} & $4132.058$ & $1.61$ & $-0.675$ & $ 20$ & $  2.63$ & $  2.86$ \\
\ion{Fe}{1} & $4143.868$ & $1.56$ & $-0.511$ & $ 32$ & $  2.68$ & $  2.91$ \\
\ion{Fe}{1} & $4202.029$ & $1.49$ & $-0.708$ & $ 28$ & $  2.71$ & $  2.94$ \\
\ion{Fe}{1} & $4216.184$ & $0.00$ & $-3.356$ & $  9$ & $  3.03$ & $  3.32$ \\
\ion{Fe}{1} & $4250.787$ & $1.56$ & $-0.714$ & $ 24$ & $  2.70$ & $  2.93$ \\
\ion{Fe}{1} & $4260.474$ & $2.40$ & $ 0.109$ & $ 16$ & $  2.63$ & $  2.82$ \\
\ion{Fe}{1} & $4271.761$ & $1.49$ & $-0.164$ & $ 53$ & $  2.64$ & $  2.87$ \\
\ion{Fe}{1} & $4294.125$ & $1.49$ & $-1.110$ & $ 20$ & $  2.91$ & $  3.14$ \\
\ion{Fe}{1} & $4325.762$ & $1.61$ & $ 0.006$ & $ 51$ & $  2.58$ & $  2.81$ \\
\ion{Fe}{1} & $4375.930$ & $0.00$ & $-3.031$ & $ 17$ & $  3.01$ & $  3.30$ \\
\ion{Fe}{1} & $4383.545$ & $1.49$ & $ 0.200$ & $ 67$ & $  2.55$ & $  2.78$ \\
\ion{Fe}{1} & $4404.750$ & $1.56$ & $-0.142$ & $ 46$ & $  2.57$ & $  2.80$ \\
\ion{Fe}{1} & $4415.122$ & $1.61$ & $-0.615$ & $ 27$ & $  2.72$ & $  2.95$ \\
\ion{Fe}{1} & $4427.310$ & $0.05$ & $-2.924$ & $ 13$ & $  2.80$ & $  3.10$ \\
\ion{Fe}{1} & $4920.503$ & $2.83$ & $ 0.068$ & $  6$ & $  2.67$ & $  2.85$ \\
\ion{Fe}{1} & $5232.940$ & $2.94$ & $-0.058$ & $  5$ & $  2.77$ & $  2.95$ \\
\ion{Fe}{1} & $5269.537$ & $0.86$ & $-1.321$ & $ 49$ & $  2.89$ & $  3.15$ \\
\ion{Fe}{1} & $5328.039$ & $0.92$ & $-1.466$ & $ 35$ & $  2.84$ & $  3.10$ \\
\ion{Fe}{1} & $5371.490$ & $0.96$ & $-1.645$ & $ 25$ & $  2.87$ & $  3.13$ \\
\ion{Fe}{1} & $5397.128$ & $0.92$ & $-1.993$ & $ 14$ & $  2.87$ & $  3.13$ \\
\ion{Fe}{1} & $5405.775$ & $0.99$ & $-1.844$ & $ 19$ & $  2.95$ & $  3.20$ \\
\ion{Fe}{1} & $5429.697$ & $0.96$ & $-1.879$ & $ 17$ & $  2.88$ & $  3.14$ \\
\ion{Fe}{1} & $5434.524$ & $1.01$ & $-2.122$ & $  8$ & $  2.79$ & $  3.05$ \\
\ion{Fe}{1} & $5446.917$ & $0.99$ & $-1.914$ & $ 14$ & $  2.85$ & $  3.10$ \\
\ion{Fe}{1} & $5455.610$ & $1.01$ & $-2.091$ & $ 11$ & $  2.94$ & $  3.20$ \\
\ion{Fe}{2} & $5018.440$ & $2.89$ & $-1.220$ & $  4$ & $  2.53$ & $  2.57$ \\
\ion{Fe}{2} & $5169.033$ & $2.89$ & $-1.303$ & synth & $  2.86$ & $  2.90$ \\
\ion{Ni}{1} & $3414.760$ & $0.03$ & $-0.014$ & $ 66$ & $  1.05$ & $  1.35$ \\
\ion{Ni}{1} & $3423.704$ & $0.21$ & $-0.760$ & $ 30$ & $  1.16$ & $  1.45$ \\
\ion{Ni}{1} & $3433.554$ & $0.03$ & $-0.668$ & $ 47$ & $  1.22$ & $  1.52$ \\
\ion{Ni}{1} & $3452.885$ & $0.11$ & $-0.910$ & $ 35$ & $  1.32$ & $  1.61$ \\
\ion{Ni}{1} & $3610.461$ & $0.11$ & $-1.149$ & $ 28$ & $  1.38$ & $  1.67$ \\
\ion{Ni}{1} & $3783.524$ & $0.42$ & $-1.310$ & $ 10$ & $  1.32$ & $  1.59$ \\
\ion{Ni}{1} & $3807.138$ & $0.42$ & $-1.205$ & $ 12$ & $  1.28$ & $  1.55$ \\
\ion{Ni}{1} & $3858.292$ & $0.42$ & $-0.936$ & $ 21$ & $  1.30$ & $  1.56$ \\
\ion{Sr}{2} & $4077.709$ & $0.00$ & $ 0.167$ & $<10$ & $<-2.90$ & $<-2.75$ \\
\ion{Y }{2} & $3788.700$ & $0.10$ & $-0.070$ & $<10$ & $<-1.80$ & $<-1.64$ \\
\ion{Zr}{2} & $4211.907$ & $0.53$ & $-1.083$ & $<10$ & $< 0.11$ & $< 0.25$ \\
\ion{Ba}{2} & $4934.076$ & $0.00$ & $-0.150$ & $<10$ & $<-2.55$ & $<-2.37$ \\
\ion{Eu}{2} & $4129.725$ & $0.00$ & $ 0.173$ & $<10$ & $<-2.19$ & $<-2.03$ \\
\enddata
\end{deluxetable} 

\clearpage

\begin{deluxetable}{ll} 
\tablecolumns{2} 
\tablewidth{0pt} 
\tablecaption{\label{Tab:HE0557_Velocities} RADIAL VELOCITIES OF {\hes}.} 
\tablehead{
  \colhead{MJD} & \colhead{V$_{\rm r}$} \\
  \colhead{}    & \colhead{kms$^{-1}$}  \\
  \colhead{(1)} & \colhead{(2)} 
  }
\startdata 
 53771.029 & 211.44\\
 53771.072 & 211.67\\
 53771.159 & 211.61\\
 53791.025 & 211.98\\
 53792.035 & 211.90\\
 53792.078 & 212.02\\
 53813.999 & 211.84\\
\enddata
\end{deluxetable} 

\clearpage

\begin{deluxetable}{llllll} 
\tablecolumns{6} 
\tablewidth{0pt} 
\tablecaption{\label{Tab:HE0557_Photometry} BROAD-BAND PHOTOMETRY OF {\hen} AND {\hes}.} 
\tablehead{
  \colhead{Object} & \colhead{$V$} & \colhead{$B-V$} & \colhead{$V-R$} & \colhead{$K$}& \colhead{$V-K$}\\
  \colhead{(1)} & \colhead{(2)} & \colhead{(3)} & \colhead{(4)} & \colhead{(5)}& \colhead{(6)}
  }
\startdata 
 {\hen}              & 15.17\tablenotemark{a}   & 0.69\tablenotemark{a}  & 0.44\tablenotemark{a}   & 13.24\tablenotemark{a}  & 1.93\tablenotemark{a}  \\        
 {\hes}              & 15.454\tablenotemark{b}  & 0.712\tablenotemark{b} & 0.463\tablenotemark{b}  & 13.194\tablenotemark{c} & 2.254\tablenotemark{d} \\      
\enddata
\tablenotetext{a}{From \citet[][Table 2, ``adopted'' values]{HE0107_ApJ}}
\tablenotetext{b}{From \citet{Beersetal:2007};  errors in $V, B-V$ and $V-R$ are 0.004, 0.008, and 0.006, respectively }
\tablenotetext{c}{From 2MASS \citep{Skrutskieetal:2006};  error is 0.037 } 
\tablenotetext{d}{On the Johnson-Glass system;  error is 0.037 }
\end{deluxetable} 

\clearpage

\begin{deluxetable}{lrlcrlrc} 
\tablecolumns{8} 
\tablewidth{0pt} 
\tablecaption{\label{Tab:HE0557_Teff} DERIVATION OF  $T_{\mbox{\scriptsize eff}}$ FOR  {\hes} AND {\hen}.} 
\tablehead{
  \colhead{} & \multicolumn{2}{c}{{\hes}} & \colhead{} & \multicolumn{2}{c}{{\hen}} & \colhead{}\\
  \cline{2-3} \cline{5-6} \\ 
  \colhead{Measured\tablenotemark{a}} & \colhead{Value} & \colhead{Derived {\tefft}} & \colhead{}
  & \colhead{Value} & \colhead{Derived {\tefft}} & \colhead{$\Delta\teffm$} & \colhead{References\tablenotemark{b}}\\
  \colhead{quantity} & \colhead{}      & \colhead{(K)}              & \colhead{}
  & \colhead{} & \colhead{(K)} & \colhead{(K)} & \colhead{}\\
  \colhead{(1)} & \colhead{(2)}      & \colhead{(3)}              & \colhead{}
  & \colhead{(4)} & \colhead{(5)} & \colhead{(6)} & \colhead{(7)}
  }
\startdata
$H\alpha$           &       ...     &  4900 $\pm$ 100    & &     ...    &   5180 $\pm$  70    &  $-$280   &     1 \\

$(B-V)_{0}$/J       &     0.672     &  5170 $\pm$  50    & &  0.680     &   5150 $\pm$  50    &      20   &    2 \\
$(V-R)_{0}$/C       &     0.438     &  5140 $\pm$ 100    & &  0.434     &   5170 $\pm$ 100    &   $-$30   &    2 \\
$(V-K)_{0}$/JG      &     2.125     &  5060 $\pm$  70    & &  1.884     &   5340 $\pm$  80    &  $-$280   &    2 \\
$(B-V)_{0}$/J       &     0.672     &  5110 $\pm$  60    & &  0.680     &   5080 $\pm$  60    &      30   &    3 \\
$(V-K)_{0}$/JTCS    &     2.179     &  4920 $\pm$  70    & &  1.935     &   5230 $\pm$  80    &  $-$310   &    3 \\

\enddata

\tablenotetext{a} {Colors with qualifiers J, C, and JG refer to the Johnson,
  Cousins, and Johnson-Glass systems, respectively. JTCS refers to $V-K$
  colors, where transformation from 2MASS to Johnson-Glass was effected by
  using the transformations of \citet{Carpenter:2001} and thence to the TCS
  system following \citet{Alonsoetal:1998}}

\tablenotetext{b} {References $-$ (1) \citet{Fuhrmannetal:1993}; (2)
  \citet{Houdasheltetal:2000}; (3) \citet{Alonsoetal:1999b,Alonsoetal:2001}}

\end{deluxetable} 

\clearpage

\begin{deluxetable}{lrrrr} 
\tablecolumns{5} 
\tablewidth{0pt} 
\tablecaption{\label{Tab:StellarParameters} STELLAR PARAMETERS FOR {\hes}.} 
\tablehead{
  \colhead{} & \multicolumn{2}{c}{Spectroscopy} & \multicolumn{2}{c}{Photometry}\\ 
  \colhead{Parameter} & \colhead{Value} & \colhead{$\sigma$\tablenotemark{a}} & \colhead{Value} & \colhead{$\sigma$\tablenotemark{a}}\\
  \colhead{(1)} & \colhead{(2)} & \colhead{(3)} & \colhead{(4)} & \colhead{(5)}
  }
\startdata 
 $T_{\mbox{\scriptsize eff}}$ (K)    & $4900$    & $100$  & $5100$   &  $50$\\                 
 $\log g$ (cgs)                      & $2.2$      & $0.3$  & $2.2$    & $0.3$\\               
 $\mbox{[Fe/H]}$                     & $-4.8$     & $0.2$  & $-4.7$   & $0.2$\\              
 $\xi_{\rm micr}$ (km\,s$^{-1}$)     & $1.8$      & $0.2$  & $1.8$    & $0.2$
\enddata
\tablenotetext{a}{Internal error estimate}
\end{deluxetable} 

\clearpage

\begin{deluxetable}{lrrrrrrrrl} 
\tablecolumns{10} 
\tablewidth{0pc} 
\tablecaption{\label{Tab:1DLTEAbundances} 1D LTE ABUNDANCES OF {\hes}.} 
\tablehead{ 

\colhead{Spec-} & \colhead{$N_{\mbox{\scriptsize lines}}$} & 
\colhead{$\log\epsilon (\mbox{X})$} & \colhead{$\log\epsilon (\mbox{X})$} & 
\colhead{$\sigma_{\log\epsilon}$\tablenotemark{a}} & \colhead{3D\tablenotemark{b}} & 
\colhead{$\log\epsilon (\mbox{X})_{\odot}$\tablenotemark{c}} & 
\colhead{[X/Fe]\tablenotemark{d}} &
\colhead{[X/Fe]\tablenotemark{e}} & 
\colhead{Notes}\\

\colhead{ies} & \colhead{} & \colhead{($4900$\,K)} & 
\colhead{($5100$\,K)} & \colhead{} & \colhead{Corr.} & \colhead{} & \colhead{($4900$\,K)} & 
\colhead{($5100$\,K)} & \colhead{}\\

\colhead{(1)} & \colhead{(2)} & \colhead{(3)} & 
\colhead{(4)} & \colhead{(5)} & \colhead{(6)} & \colhead{(7)} & 
\colhead{(8)} & \colhead{(9)} & \colhead{(10)}
}
\startdata 
\ion{Li}{1}   &  1      & $< 0.70$ & $< 0.90$ &\nodata  & $-0.2$     & \nodata & \nodata & \nodata  & synthesis\\  
  C(CH)       & \nodata & $  5.29$ & $  5.76$ & $0.10$  & $\sim-1.0$ & 8.39 & $ +1.65$ & $ +2.08$  & A-X \& B-X\\ 
  N(CN)       & \nodata & $< 5.40$ & $< 5.60$ & $0.10$  & $\sim-2.2$ & 7.78 & $<+2.37$ & $<+2.53$ & B-X\\ 
  N(NH)       & \nodata & $< 4.50$ & $< 5.00$ & $0.20$  & $\sim-1.0$ & 7.78 & $<+1.47$ & $<+1.93$ & A-X\\
  \ion{O}{1}  &  1      & $< 7.00$ & $< 7.13$ &\nodata  & \nodata    & 8.66 & $<+3.09$ & $<+3.18$ & \\
  \ion{Na}{1} &  1      & $  1.26$ & $  1.44$ &\nodata  & $-0.1$     & 6.17 & $ -0.16$ & $ -0.02$ & \\
  \ion{Mg}{1} &  5      & $  3.03$ & $  3.20$ & $0.07$  & $-0.1$     & 7.53 & $ +0.25$ & $ +0.38$ & \\
  \ion{Al}{1} &  1      & $  1.01$ & $  1.22$ &\nodata  & $-0.1$     & 6.37 & $ -0.61$ & $ -0.44$ & \\
  \ion{Ca}{1} &  1      & $  1.81$ & $  2.03$ &\nodata  & $-0.2$     & 6.31 & $ +0.25$ & $ +0.43$ & \\
  \ion{Ca}{2} &  2      & $  2.18$ & $  2.32$ &\nodata  & $-0.2$     & 6.31 & $ +0.62$ & $ +0.72$ & \\
  \ion{Sc}{2} &  4      & $ -1.39$ & $ -1.22$ & $0.11$  & $-0.1$     & 3.05 & $ +0.31$ & $ +0.44$ & \\
  \ion{Ti}{2} &  5      & $  0.50$ & $  0.62$ & $0.10$  & $0.0$      & 4.90 & $ +0.35$ & $ +0.43$ & \\
  \ion{Cr}{1} &  2      & $  0.19$ & $  0.44$ &\nodata  & $-0.3$     & 5.64 & $ -0.70$ & $ -0.49$ & \\
  \ion{Mn}{1} &  1      & $< 0.01$ & $ <0.29$ &\nodata  & $-0.4$     & 5.39 & $<-0.63$ & $<-0.39$ & \\
  \ion{Fe}{1} & 60      & $  2.77$ & $  3.02$ & $0.17$  & $-0.2$     & 7.45 & $ +0.07$ & $ +0.28$ & \\
  \ion{Fe}{2} &  2      & $  2.70$ & $  2.74$ &\nodata  & $+0.1$     & 7.45 & \nodata  & \nodata  & \\
  \ion{Co}{1} &  2      & $  0.20$ & $  0.50$ &\nodata  & $-0.3$     & 4.92 & $ +0.04$ & $ +0.30$ & co-addition\\
  \ion{Ni}{1} &  8      & $  1.26$ & $  1.54$ & $0.10$  & $-0.3$     & 6.23 & $ -0.22$ & $ +0.02$ & \\
  \ion{Sr}{2} &  1      & $<-2.90$ & $<-2.75$ &\nodata  & $-0.2$     & 2.92 & $<-1.07$ & $<-0.98$ & \\
  \ion{Y}{2}  &  1      & $<-1.80$ & $<-1.64$ &\nodata  & \nodata    & 2.21 & $<+0.74$ & $<+0.86$ & \\
  \ion{Zr}{2} &  1      & $< 0.11$ & $< 0.25$ &\nodata  & \nodata    & 2.59 & $<+2.27$ & $<+2.37$ & \\
  \ion{Ba}{2} &  1      & $<-2.55$ & $<-2.37$ &\nodata  & $-0.3$     & 2.17 & $<+0.03$ & $<+0.17$ & \\
  \ion{Eu}{2} &  1      & $<-2.19$ & $<-2.03$ &\nodata  & $-0.5$     & 0.52 & $<+2.04$ & $<+2.16$ & \\
\enddata 
\tablenotetext{a}{Uncertainty of the fit in the case of spectrum synthesis; standard
  deviation of the mean for abundances determined from equivalent width
  measurements of $N > 2$ lines}
\tablenotetext{b}{3D Corr. = $\log\epsilon (\mbox{X})_{\rm 3D}$ -- $\log\epsilon (\mbox{X})_{\rm 1D}$, following \citet{Colletetal:2006} for {\hen}}
\tablenotetext{c}{From Asplund et al.\ (2005)}
\tablenotetext{d}{Adopting an Fe abundance of ${\rm [Fe/H]}=-4.75$, as derived
  from \ion{Fe}{2}}
\tablenotetext{e}{Adopting an Fe abundance of ${\rm [Fe/H]}=-4.71$, as derived
  from \ion{Fe}{2}}
\end{deluxetable} 

\clearpage
\begin{deluxetable}{lcccc}
\tablecolumns{5}
\tablewidth{0pt}
\tablecaption{\label{Tab:calcium} ABUNDANCE CHANGES AND $\log \epsilon$ (Ca) FOR {\tefft}~=~4900\,K, FOLLOWING MASHONKINA ET AL. (2007).}
\tablehead{
\colhead{Line} & 
\colhead{$\log \epsilon$ (non-LTE)} &
\colhead{$\log \epsilon$ ($\log gf$)} & 
\colhead{$\log \epsilon$ (solar fit)} & 
\colhead{$\log \epsilon$ (Ca)$_{\rm Diff}^{\rm non-LTE}$}\\
\colhead{} & 
\colhead{$-\log \epsilon$ (LTE)} &
\colhead{$-\log \epsilon$ (LTE)} &
\colhead{$-\log \epsilon$ (LTE)} &
\colhead{(4900\,K)}\\
\colhead{(1)} & 
\colhead{(2)} &
\colhead{(3)} & 
\colhead{(4)} & 
\colhead{(5)}
  }
\startdata
Ca I  4226 & +0.26 & +0.02\tablenotemark{a} & +0.10                    & 2.19\\
Ca II 3706 & +0.17 & --0.06\tablenotemark{b} & \nodata\tablenotemark{c} & 2.22\\
Ca II 3933 & 0.00  & --0.02\tablenotemark{b} & \nodata\tablenotemark{c} & 2.24\\
\enddata
\tablenotetext{a}{From \citet{Smith/Gallagher:1966}}
\tablenotetext{b}{From the Opacity Project \citep{Seatonetal:1994}}
\tablenotetext{c}{No solar fitting attempted due to strong blending}
\end{deluxetable}

\clearpage

\appendix

\section{Rejected lines}

In Table~\ref{Tab:RejectedLines} we list absorption lines that were excluded
from the abundance analysis for various reasons. In the case of rejections
based on visual inspection of a line in the VLT/UVES spectrum, it was rejected
when either N.C. or J.E.N. (or both) suspected it to be blended.

\begin{deluxetable}{lllrl} 
\tablecolumns{5} 
\tablewidth{0pt} 
\tablecaption{\label{Tab:RejectedLines} REJECTED ABSORPTION LINES.}
\tablehead{
  \colhead{} & \colhead{$\lambda$} & \colhead{$\chi$} & \colhead{$\log gf$} &
  \colhead{}\\
  \colhead{Species} & \colhead{({\AA})}  & \colhead{(eV)} & \colhead{(dex)} &
  \colhead{Reason for rejection}\\
  \colhead{(1)} & \colhead{(2)}  & \colhead{(3)} & \colhead{(4)} &
  \colhead{(5)}
  }
\startdata
\ion{Na}{1} & $5895.924$ & $0.000$ & $-0.184$ & Blended\tablenotemark{a}\\
\ion{Si}{1} & $3905.523$ & $1.909$ & $-1.090$ & Blended\tablenotemark{a}\\
\ion{Sc}{2} & $3613.829$ & $0.022$ & $ 0.416$ & Blended\tablenotemark{a}\\
\ion{Sc}{2} & $3630.742$ & $0.008$ & $ 0.220$ & Blended\tablenotemark{a}\\
\ion{Ti}{2} & $3444.314$ & $0.151$ & $-0.810$ & Blended\tablenotemark{a}\\
\ion{Ti}{2} & $3477.187$ & $0.122$ & $-0.967$ & Blended\tablenotemark{a}\\
\ion{Ti}{2} & $3491.066$ & $0.113$ & $-1.060$ & Blended\tablenotemark{a}\\
\ion{Ti}{2} & $3641.331$ & $1.237$ & $-0.710$ & Blended\tablenotemark{a}\\
\ion{Ti}{2} & $3685.19 $ & $0.574$ & $-0.040$ & Affected by Balmer line\\
\ion{Ti}{2} & $3814.584$ & $0.574$ & $-1.700$ & Blended\tablenotemark{a}\\
\ion{Fe}{1} & $3440.606$ & $0.000$ & $-0.673$ & low $S/N$\\
\ion{Fe}{1} & $3440.989$ & $0.052$ & $-0.958$ & low $S/N$\\
\ion{Fe}{1} & $3443.877$ & $0.087$ & $-1.374$ & low $S/N$\\
\ion{Fe}{1} & $3475.450$ & $0.087$ & $-1.054$ & low $S/N$\\
\ion{Fe}{1} & $3476.702$ & $0.121$ & $-1.507$ & low $S/N$\\
\ion{Fe}{1} & $3490.574$ & $0.052$ & $-1.105$ & low $S/N$\\
\ion{Fe}{1} & $3497.841$ & $0.110$ & $-1.549$ & low $S/N$\\
\ion{Fe}{1} & $3521.261$ & $0.915$ & $-0.988$ & low $S/N$\\
\ion{Fe}{1} & $3558.515$ & $0.990$ & $-0.629$ & low $S/N$\\
\ion{Fe}{1} & $3565.379$ & $0.958$ & $-0.133$ & low $S/N$\\
\ion{Fe}{1} & $3581.193$ & $0.859$ & $ 0.406$ & low $S/N$\\
\ion{Fe}{1} & $3585.705$ & $0.915$ & $-1.187$ & low $S/N$\\
\ion{Fe}{1} & $3586.985$ & $0.990$ & $-0.796$ & low $S/N$\\
\ion{Fe}{1} & $3608.859$ & $1.011$ & $-0.100$ & low $S/N$\\
\ion{Fe}{1} & $3618.768$ & $0.990$ & $-0.003$ & low $S/N$\\
\ion{Fe}{1} & $3631.463$ & $0.958$ & $-0.036$ & low $S/N$\\
\ion{Fe}{1} & $3647.843$ & $0.915$ & $-0.194$ & low $S/N$\\
\ion{Fe}{1} & $3705.57 $ & $0.052$ & $-1.334$ & Affected by Balmer line\\
\ion{Fe}{1} & $3709.25 $ & $0.915$ & $-0.646$ & Affected by Balmer line\\
\ion{Fe}{1} & $3795.002$ & $0.990$ & $-0.761$ & Blended\tablenotemark{a}\\
\ion{Fe}{1} & $3840.44 $ & $0.990$ & $-0.506$ & Affected by Balmer line\\
\ion{Fe}{1} & $3872.50 $ & $0.991$ & $-0.910$ & Blended\tablenotemark{a}\\
\ion{Fe}{1} & $4238.02 $ & $3.417$ & $-1.286$ & Blended with a CH line\\
\ion{Co}{1} & $3873.114$ & $0.432$ & $-0.660$ & Blended with a CH line\\
\ion{Ni}{1} & $3413.472$ & $0.165$ & $-1.480$ & Blended\tablenotemark{a}\\
\ion{Ni}{1} & $3437.275$ & $0.000$ & $-1.192$ & Blended\tablenotemark{a}\\
\ion{Ni}{1} & $3472.542$ & $0.109$ & $-0.810$ & Blended\tablenotemark{a}\\
\ion{Ni}{1} & $3483.770$ & $0.275$ & $-1.110$ & Blended\tablenotemark{a}\\
\ion{Ni}{1} & $3492.954$ & $0.109$ & $-0.250$ & Blended\tablenotemark{a}\\
\ion{Ni}{1} & $3500.846$ & $0.165$ & $-1.279$ & Blended\tablenotemark{a}\\
\ion{Ni}{1} & $3515.049$ & $0.109$ & $-0.211$ & Blended\tablenotemark{a}\\
\ion{Ni}{1} & $3524.535$ & $0.025$ & $ 0.008$ & Blended\tablenotemark{a}\\
\ion{Ni}{1} & $3597.700$ & $0.212$ & $-1.100$ & Blended\tablenotemark{a}\\
\ion{Ni}{1} & $3775.565$ & $0.423$ & $-1.393$ & Affected by Balmer line\\
\ion{Ni}{1} & $4231.027$ & $3.542$ & $-1.415$ & Blended with a CH line\\
\enddata
\tablenotetext{a}{Based on visual inspection of the VLT/UVES spectrum}
\end{deluxetable} 


\clearpage

\begin{figure*}[htbp]
\begin{center}
\epsscale{.65}
\plotone{f1.eps}
\end{center}
\centering

\caption{\label{Fig:lowfe_MDF} Histogram of [Fe/H] for stars having
  high-resolution, high $S/N$, abundance analyses and [Fe/H] $<$
  --3.0, from the work of
  \citet{Aokietal:2002d,Aokietal:2004a,Aokietal:2006},
  \citet{KeckpaperII}, \citet{Cayreletal:2004}, \citet{KeckpaperIV},
  \citet{HE0107_Nature,HE0107_ApJ},
  \citet{Frebeletal:2005,Frebeletal:2007a},
  \citet{McWilliametal:1995b},
  \citet{Norrisetal:1997b,Norrisetal:2000,Norrisetal:2001},
  \citet{Plez/Cohen:2005}, and \citet{Ryanetal:1991,
  Ryanetal:1996}. Note that the proportion of carbon-rich objects
  increases dramatically as one moves to lower abundances.}

\end{figure*}

\clearpage

\begin{figure*}[htbp]
  \begin{center}
  \epsscale{.65}
  \plotone{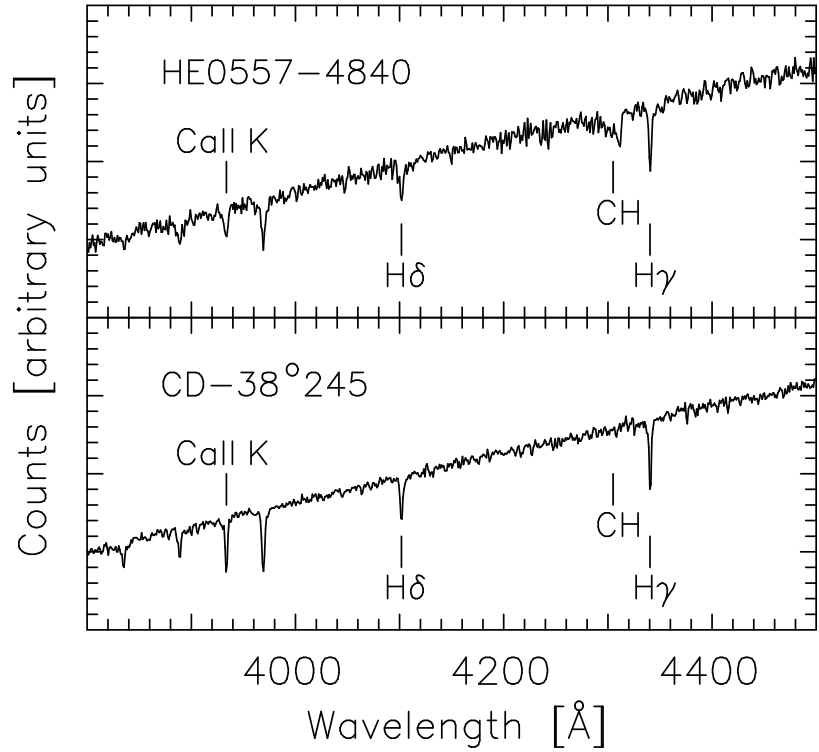}
  \end{center}
  \centering

 \caption{\label{Fig:HE0557_lowres} Medium resolution spectra of
  {\hes} and {\cd} ([Fe/H]~=~--4.0). Note that while the hydrogen
  lines have similar strengths in the two objects, the Ca II K line is
  weaker in {\hes}, suggestive of a lower Ca abundance. Note also,
  however, that the G-band of CH is considerably stronger in {\hes}.}

\end{figure*}

\clearpage

\begin{figure*}[htbp]
  \begin{center}
  \plotone{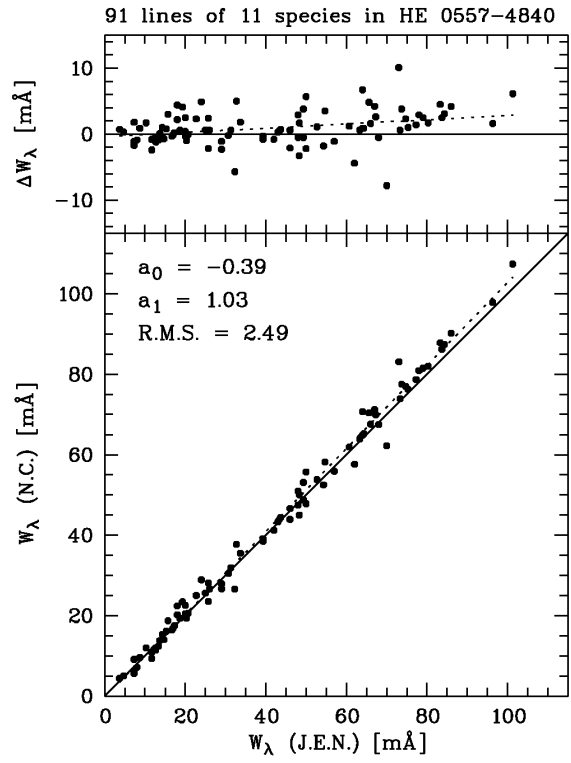}
  \end{center}
  \centering

 \caption{\label{Fig:EqwTestHE0557} Comparison of equivalent width
    measured by N.C. and J.E.N. in the VLT/UVES spectrum of {\hes}.}

\end{figure*}

\clearpage

\begin{figure*}[htbp]
  \begin{center}
  \epsscale{1.00}
  \plotone{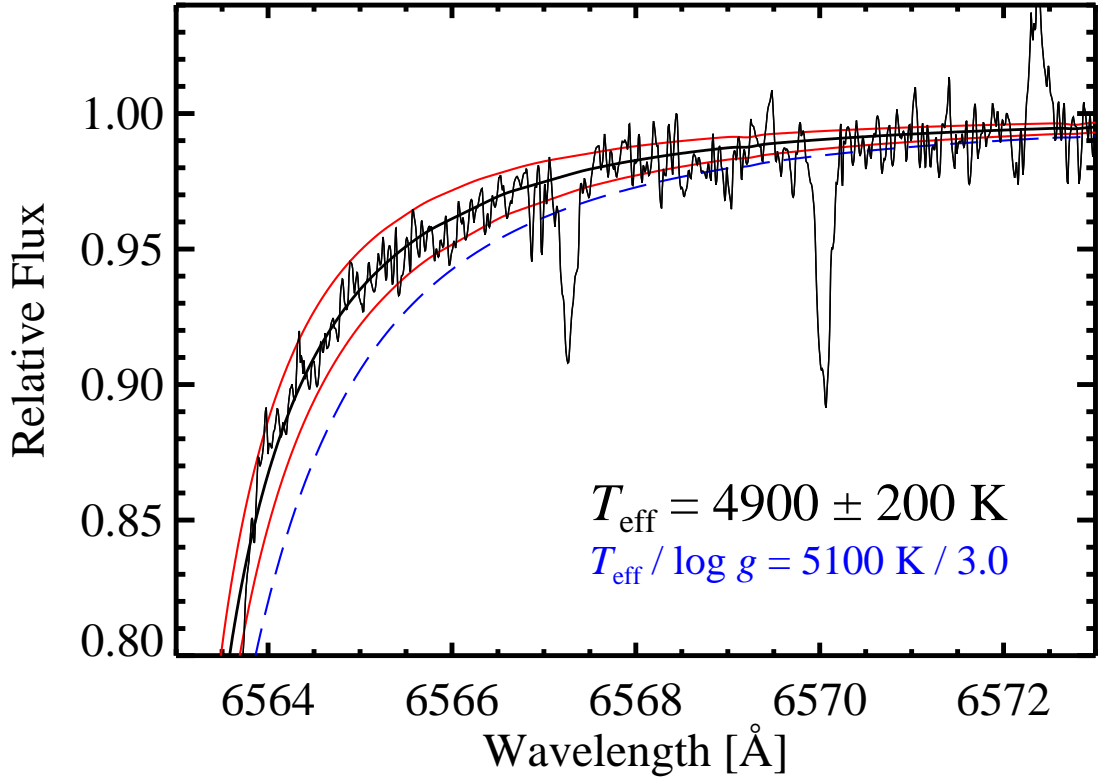}
  \vspace{-1cm}
  \end{center}
  \centering

 \caption{\label{Fig:Halpha} Comparison between the observed red wing
  of H$\alpha$ of \hes\ and synthetic spectra for {\tefft}\,=\,4700\,K
  (upper continuous line, in red), 4900\,K (middle line, in black) and
  5100\,K (lower line, in red), top to bottom (all with log$g$ =
  2.2). A synthetic profile for {\tefft}\,=\,5100\,K and log
  $g$\,=\,3.0 is also shown (dashed line, in blue, see
  \S\ref{logg}). At the given signal-to-noise ratio, an effective
  temperature of 4900 $\pm$ 100\,K ($1\sigma$) is indicated. }

\end{figure*}

\clearpage

\begin{figure*}[htbp]
  \begin{center}
  \epsscale{1.00}
  \plotone{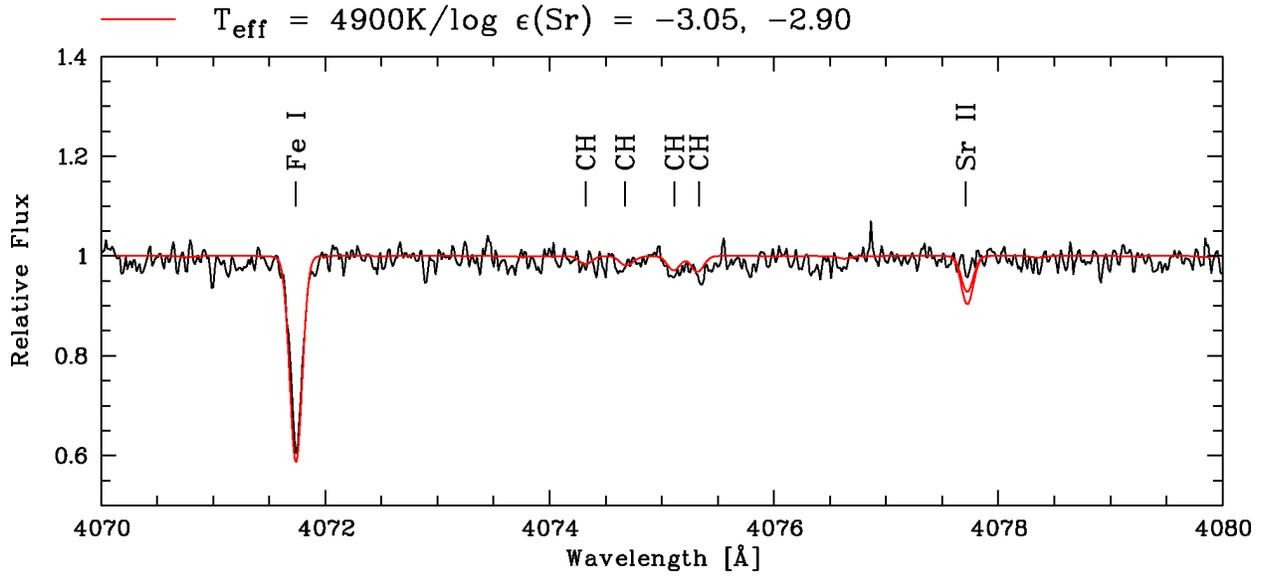}
  \end{center}
  \centering

  \caption{\label{Fig:SrII4077synth} Spectrum synthesis of the Sr II
 4077.71 line used to confirm the upper limit for the abundance of
 strontium. (The synthetic spectra (smooth red lines) have been
 broadened with a gaussian profile having FWHM = 8.5 km\,s$^{-1}$,
 corresponding to R = 35,000.)}
 
\end{figure*}

\clearpage

\begin{figure*}[htbp]
  \begin{center}
  \epsscale{.55}
  \plotone{f6.eps}
  \end{center}
  \centering
 
 \caption{\label{Fig:CoaddCo} Comparison of observed and synthetic
  co-added spectra for the strongest two lines \ion{Co}{1} lines in
  the interval 3400--3600\,{\AA} in {\cd} and {\hes} (for
  $\teffm=4900$\,K).  The lines co-added are identified, and the Co
  abundances given, in the figure. (The synthetic spectra (smooth red
  lines) for {\cd} and {\hes} have been convolved with gaussian
  profiles having FWHM = 11.0 and 8.5 km\,s$^{-1}$, respectively,
  appropriate to the instrumental setups.) See text for details.}

\end{figure*}

\clearpage

\begin{figure*}[htbp]
  \begin{center}
  \epsscale{1.00}
  \plotone{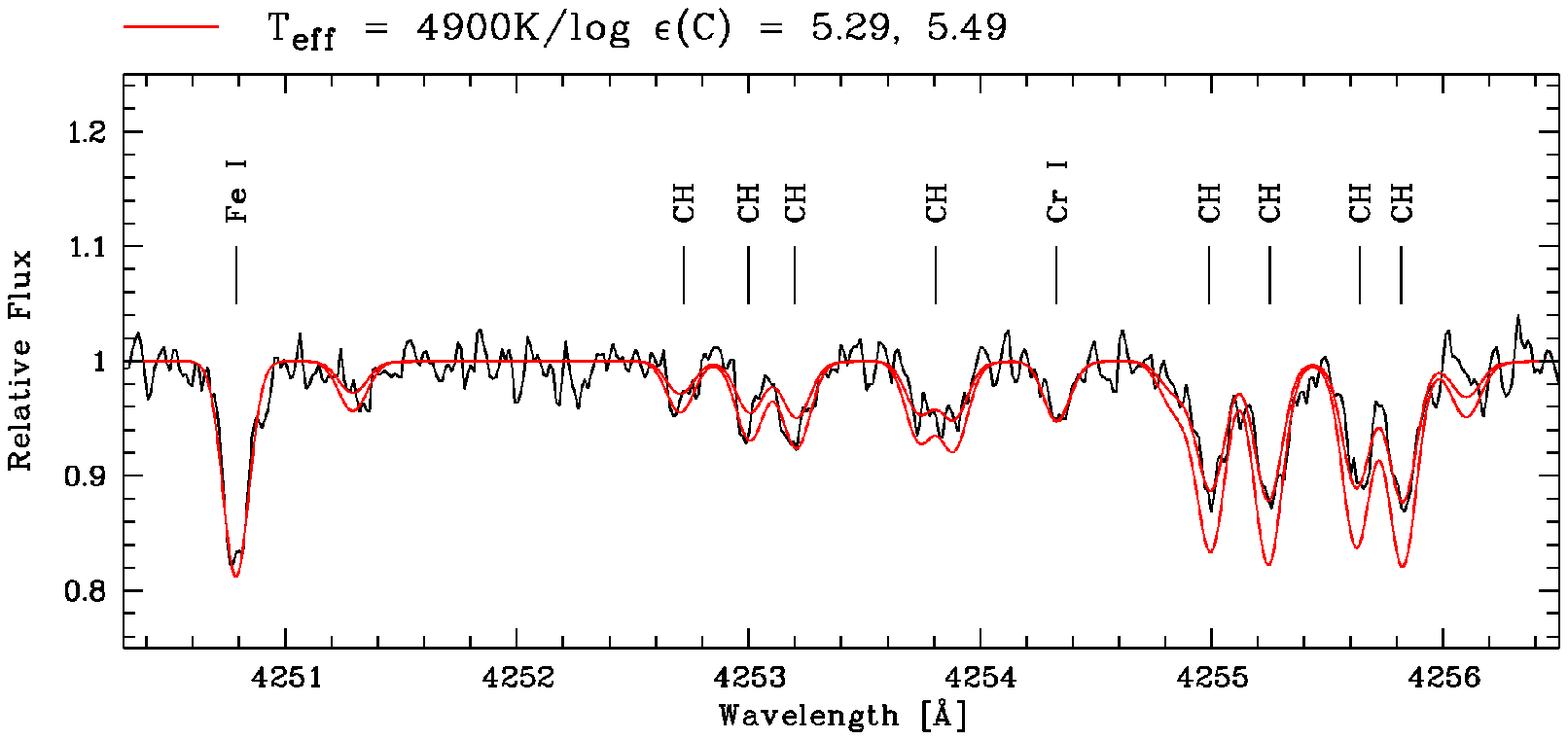}
  \end{center}
  \centering

\caption{\label{Fig:CH4250fit} Spectrum synthesis (smooth red lines)
        of CH A-X lines in {\hes} for $\teffm=4900$\,K and carbon
        abundances of $\log\epsilon\left({\rm C}\right)=5.29$\,dex
        (best fit) and $5.49$\,dex.}
 
\end{figure*}

\clearpage

\begin{figure*}[htbp]
  \begin{center}
  \epsscale{1.00}
  \plotone{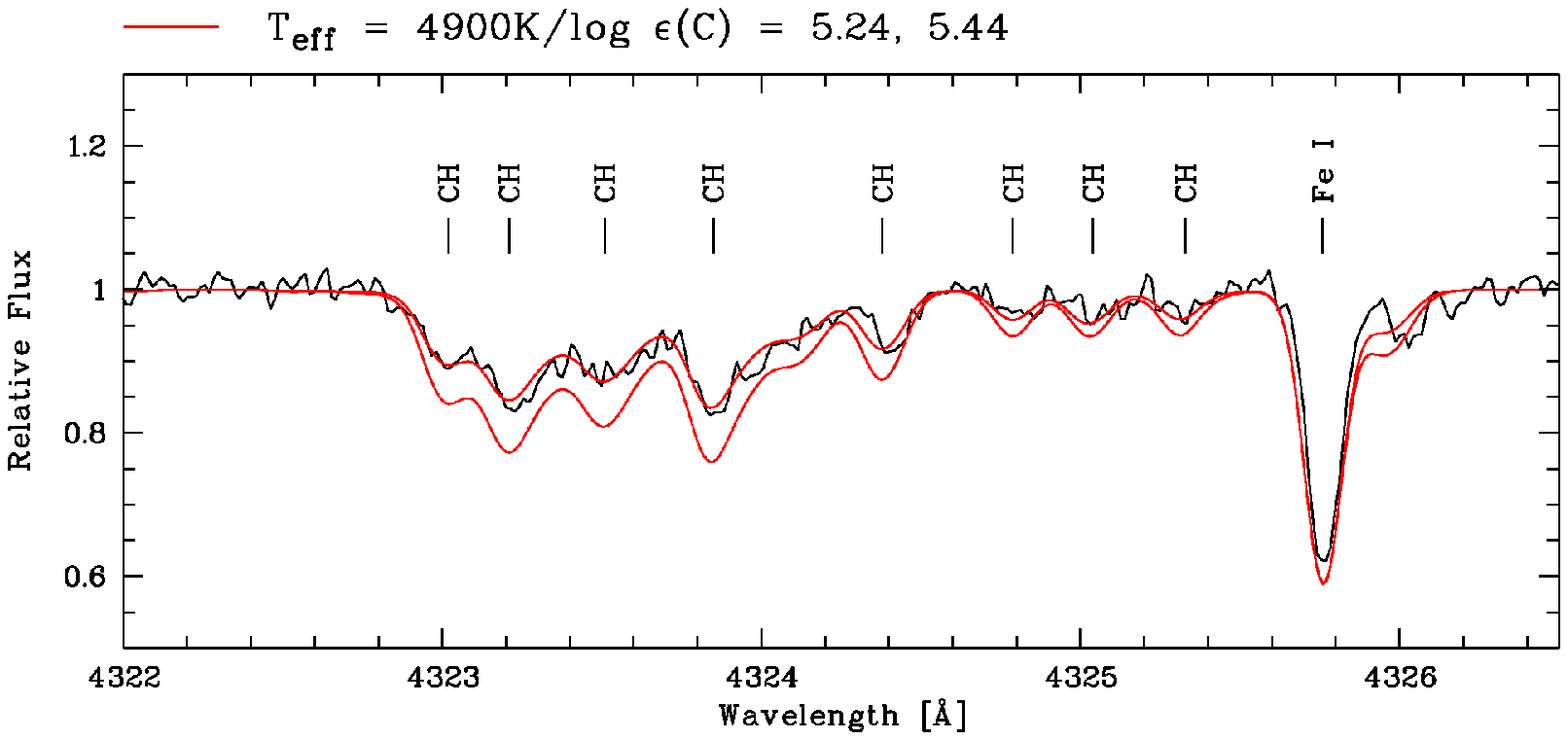}
  \end{center}
  \centering
 
\caption{\label{Fig:CH4323fit} Spectrum synthesis (smooth red lines)
  of CH A-X lines in {\hes} for $\teffm=4900$\,K and carbon abundances
  of $\log\epsilon\left({\rm C}\right)=5.24$\,dex (best fit) and
  $5.44$\,dex.}

\end{figure*}

\clearpage

\begin{figure*}[htbp]
  \begin{center}
  \epsscale{1.00}
  \plotone{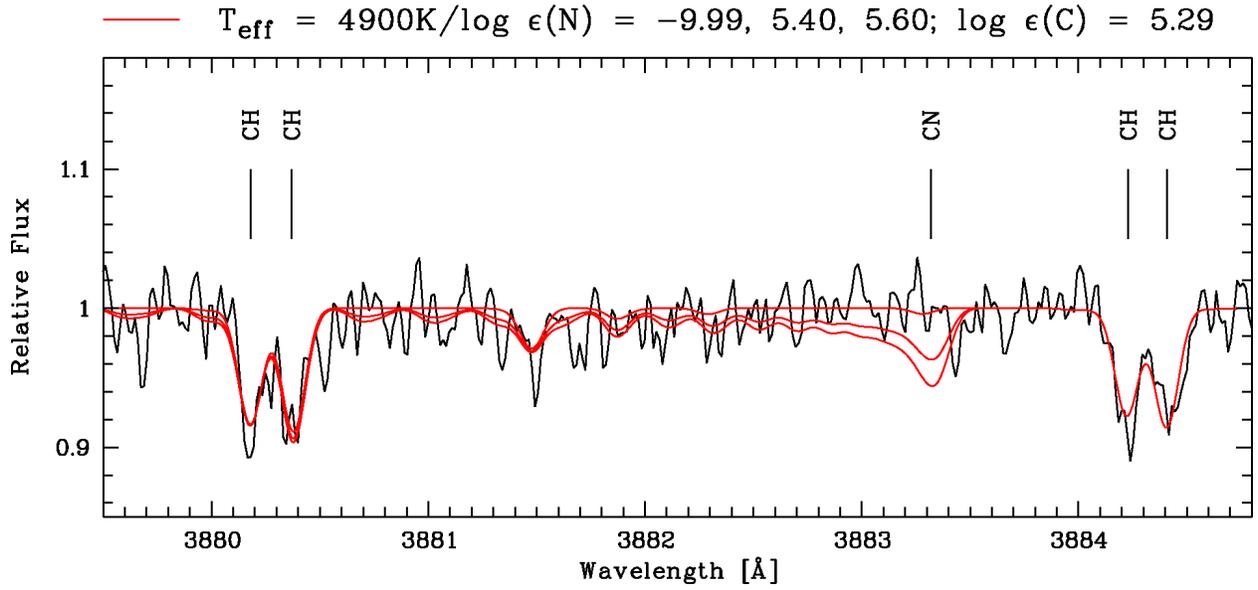}
  \end{center}
  \centering
 
\caption{\label{Fig:CN3883synth} Spectrum synthesis (smooth red lines)
    of the $(0,0)$ band head of the violet CN system in {\hes} for
    $\teffm=4900$\,K, $\log\epsilon\left({\rm C}\right)=5.29$\, and
    nitrogen abundances of $\log\epsilon\left({\rm N}\right)=-9.99$\,
    (i.e., no nitrogen), $5.40$\, (adopted upper limit) and $5.60$\,.}

\end{figure*}

\clearpage

\begin{figure*}[htbp]
  \begin{center}
  \epsscale{1.00}
  \plotone{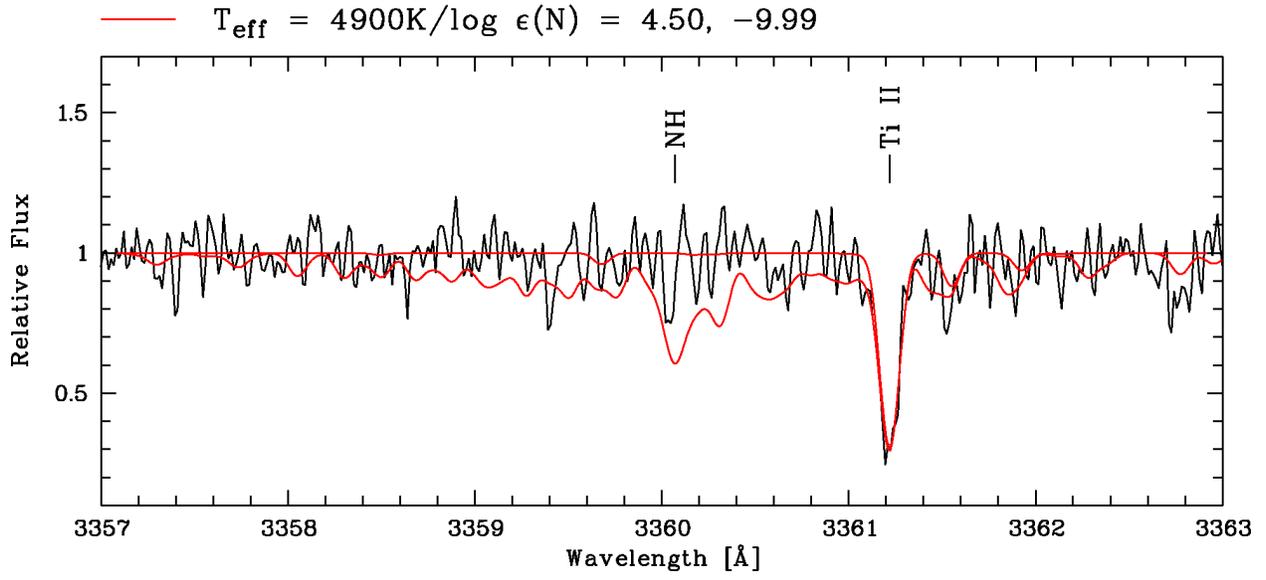}
  \end{center}
  \centering
  
\caption{\label{Fig:NHsynth} Spectrum synthesis (smooth red lines) of
    the NH band head for $\teffm=4900$\,K and nitrogen abundances of
    $\log\epsilon\left({\rm N}\right)=-9.99$\, (i.e., no nitrogen) and
    $4.50$\, (adopted upper limit).}
 
\end{figure*}
\clearpage

\begin{figure*}[htbp]
\vspace{-6mm}
  \begin{center}
  \epsscale{0.80}
  \plotone{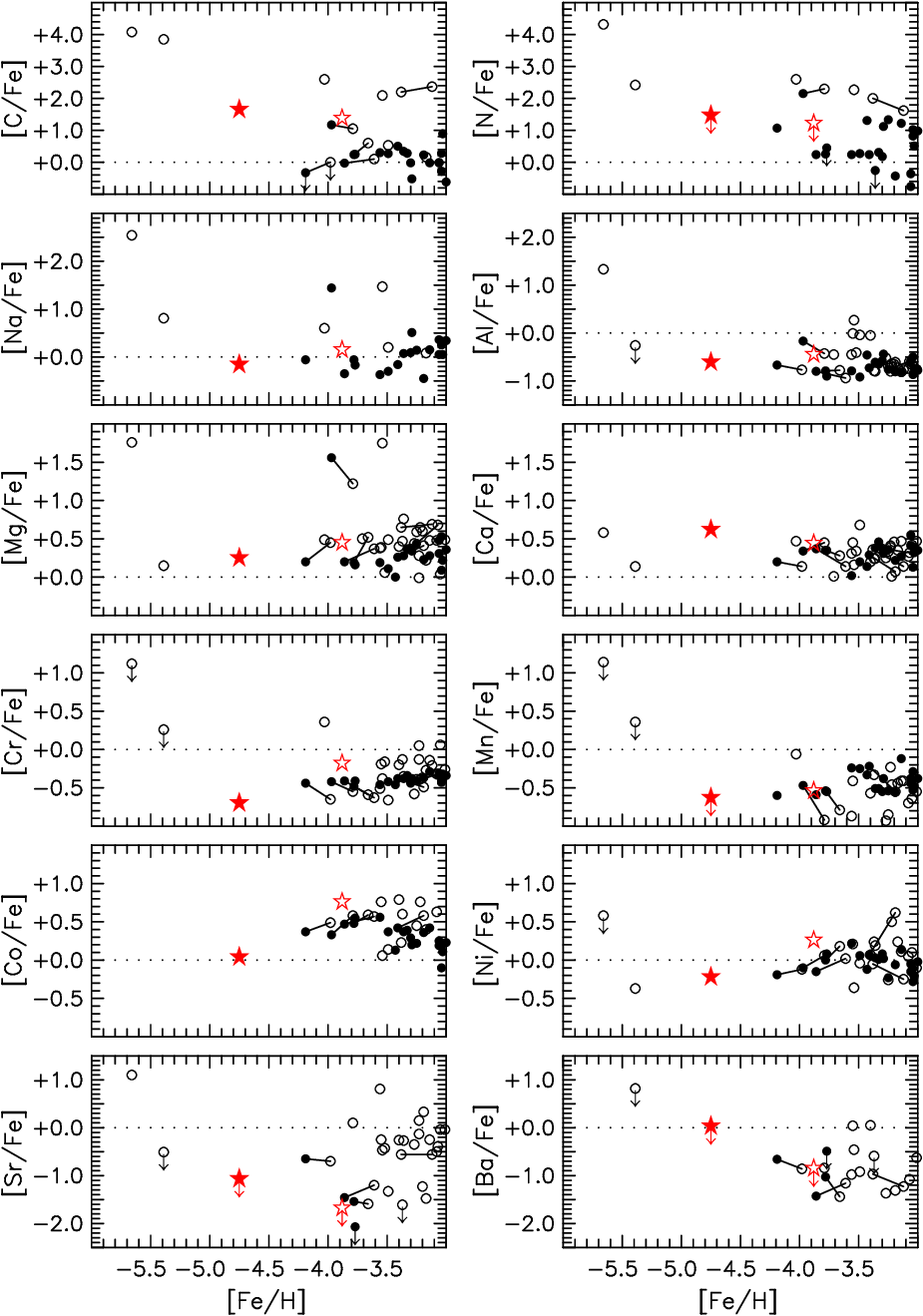}
  \end{center}
  \centering
  \vspace{-6mm}

\caption{\label{Fig:Relative_Abundances} [X/Fe] as a function of
    [Fe/H].  {\hes} is represented by a filled star while HE~1300+0157
    \citep{Frebeletal:2007a} is shown by an open one. Filled circles
    represent results of \citet{Cayreletal:2004} and
    \citet{Francoisetal:2003}, open ones those of
    \citet{Aokietal:2002d,Aokietal:2004a,Aokietal:2006},
    \citet{KeckpaperII}, \citet{KeckpaperIV},
    \citet{McWilliametal:1995b},
    \citet{Norrisetal:1997b,Norrisetal:2000,Norrisetal:2001,Norrisetal:2002},
    \citet{Plez/Cohen:2005}, and \citet{Ryanetal:1991,Ryanetal:1996}.}

\end{figure*}

\end{document}